\documentclass[prd,amsfonts,onecolumn,superscriptaddress,aps,nofootinbib,11pt]{revtex4}

\pdfoutput=1
\usepackage[utf8]{inputenc}
\usepackage{amsmath,amssymb,mathrsfs}
\usepackage{graphicx}
\usepackage{enumerate}
\usepackage{multirow}
\usepackage{float}
\usepackage[usenames,dvipsnames]{xcolor} 
\usepackage{hyperref}
\usepackage{slashed}
\usepackage{soul}
\hypersetup{colorlinks=true,
		breaklinks=true,
    linkcolor=Blue,          
    citecolor=Plum,        
    urlcolor=YellowOrange           
}

\newcommand{\met}{\not\!\! E_T}
\newcommand{\mllvv}{M_{\ell\ell\nu\nu}}

\begin{document}
\title{Reconstruction of Missing Resonances Combining \\
Nearest Neighbors Regressors and Neural Network Classifiers}

\author{Alexandre Alves}
\email{aalves@unifesp.br}
\affiliation{Departamento de Física, Universidade Federal de São Paulo, Diadema, 09913-030, Brazil}
\affiliation{Instituto de Física, Universidade de São Paulo, \\
R. do Matão 1371, 05508-090 São Paulo, Brazil}
\author{C. H. Yamaguchi}
\email{carlos.yamaguchi@usp.br}
\affiliation{Instituto de Física, Universidade de São Paulo, \\
R. do Matão 1371, 05508-090 São Paulo, Brazil}


\begin{abstract}
 Neutrinos, dark matter, and long-lived neutral particles traverse the particle detectors unnoticed, carrying away information about their parent particles and interaction sources needed to reconstruct key variables like resonance peaks in invariant mass distributions. In this work, we show that a $k$-nearest neighbors regressor algorithm combined with deep neural network classifiers, a $k$NN, is able to accurately recover binned distributions of the fully leptonic $WW$ mass of a new heavy Higgs boson and its Standard Model backgrounds from the observable detector level information at disposal. The output of the regressor can be used to train even stronger classifiers to separate signals and backgrounds in the fully leptonic case and guarantee the selection of on-mass-shell Higgs bosons with enhanced statistical significance. The method assumes previous knowledge of the event classes and model parameters, thus suitable for post-discovery studies. 
\end{abstract}

\maketitle
\section{Introduction}

 Uncovering the nature of dark matter and neutrinos will undoubtedly reveal a deeper structure of fundamental physics. If dark matter exists, it should permeate the universe and fly by our detection devices, just like the neutrinos do; however, detecting them already proved to be a challenging task. Perhaps, a better idea is to produce them in large colliders and design detectors to infer their proprieties to get clues about the underlying structure of the physical laws.
 
  Multi-purpose detectors, like \href{https://atlas.cern/}{ATLAS}~\cite{doi:10.1146/annurev-nucl-101917-021038} and \href{https://cms.cern/}{CMS}~\cite{doi:10.1146/annurev-nucl-101917-021038}, can accurately detect many types of particles like photons, electrons, muons, and hadrons, but not neutral weakly interacting particles, like neutrinos and dark matter. This fact poses a problem to the particular quest for new physics manifesting as dark states. The escape of neutrinos out of the detectors prevents us from performing some key observations that could benefit from low backgrounds. For example, the Higgs boson mass and width could be even more accurately measured if the information from fully leptonic $WW,ZZ\to \ell^+\ell^{\prime -}\nu_\ell \bar{\nu}_{\ell^\prime}$, $\ell(\ell^\prime)=e,\mu$, modes were recoverable. Instead, apart from $ZZ\to 4\ell$, we need to rely upon the semi-leptonic or fully hadronic modes to perform those measurements with a significantly higher level of backgrounds. Identifying bumps and sharp thresholds in the invariant mass distribution of observable and dark states would also help disentangle new physics signals like heavy Higgs bosons~\cite{CMS:2019bnu}, Higgs pair production with one invisible Higgs~\cite{PhysRevD.95.035009,Alves:2019emf}, sleptons and charginos~\cite{CMS:2018eqb,ATLAS:2019lff}, and new gauge bosons decays neutrinos and/or dark matter from their associate backgrounds~\cite{ATLAS:2019isd,CMS:2010kax}, to name a few possibilities. Another important example where a fully-leptonic mode benefits from a clean environment is the measurement of the scattering angles for $W,Z$ bosons in polarization studies~\cite{CMS:2020ezf}.
 
 In processes where $N_\nu$ neutrinos are produced in the hard scattering, there are $4N_\nu$ unknowns that should be recovered to reconstruct the parent particles. The negative of the sum of the transverse momentum vector of all the observed objects in the event furnishes two constraints, despite not exactly equal the sum of neutrinos transverse momentum due to detector effects, contamination from neutrinos, and other missing particles from hadronic jets, for example. Mass constraints must provide the complementary information necessary for reconstruction. The number of mass constraints, $N_m$, is process dependent though and, in many cases, they do not suffice to recover the four momenta of the neutrinos if $4N_\nu\geq N_m+2$. Even in cases where sufficient mass constraints exist, like fully leptonic $t\bar{t}$ signals~\cite{Sonnenschein:2006ud,CMS:2017xio}, the misresconstruction of the neutrinos transverse momentum, combinatorial particle assignment, and ambiguities arising from the quadratic nature of the equations do not guarantee meaningful solutions for all events. 
 
 In a process-independent way, one approach to circumvent the impossibility of recovering the four-momenta of all the escaping particles is to design kinematic variables and methods that correlate with the lost information, for example, with the masses of the parent particles. Many such variables are smartly crafted to provide useful hints about decaying particles in many situations~\cite{Lester:1999tx,Barger:2011cb,Barr:2010zj,Konar:2010ma,Kawabata:2013fta,Cho:2008tj,Barr:2007hy,Konar:2015hea,Park:2020rol,Choi:2009hn,Matchev:2019bon}. Yet, none of them, by construction, is capable of recovering a resonance peak.
 
 Another approach could be using a regression algorithm to predict the neutrinos four-momenta or some variable of interest from the observed information. One might tackle tasks of that type by training an algorithm to parameterize a multivalued function $f: \mathbb{R}^n\rightarrow \mathbb{R}^m$, with a neural network, for example~\cite{Yang:2020slb,Li:2021cbp,Arratia:2021tsq,Renteria-Estrada:2021zrd}. Methods of density estimation~\cite{Menary:2021tjg} might also be useful\footnote{For more regression algorithms and applications, see~\cite{10.5555/2490546,hastie01statisticallearning,2019}.}. The fundamental difficulty in these cases is that essential information for the reconstruction of resonances, like masses and widths, are encoded in the signal events but not in the backgrounds, the only ones sufficiently known to permit the training of an algorithm. In other words, it is necessary to rely on supervised algorithms with previous knowledge of the parameters to train a regressor to recover a resonance peak; otherwise, there are no guarantees that regressors trained for backgrounds will generalize. Contrary to classification problems whose targets are mutually excluding categorical attributes, a regression task targets a real number representing a continuum data attribute. For this reason, it is much easier to build a weakly supervised or even an unsupervised algorithm for classification, but not for regression. 
 
  Assuming previous knowledge about signals, the most straightforward approach to reconstructing a mass variable involving escaping neutrinos is by interpolating a support set of events from simulations instead of adjusting the parameters of some universal function that should generalize from training to test datasets. Such an accurate and efficient algorithm for supervised regression is the $k$-nearest neighbors algorithm, as we will demonstrate in this work. As we argued, the caveat of this approach, like any other supervised regression algorithm, is that we need to know what type of event is produced in the collisions beforehand to select the correct support set for interpolation of the variable. Our approach takes advantage of the exquisite power of neural networks to classify the events. In principle, it is possible to identify signal events without any previous knowledge using outliers detection and unsupervised methods; however, as we discussed, without knowing the mass parameters, reconstructing a mass peak is challenging\footnote{Regression with unlabeled data is possible when the marginal distribution of the target is known~\cite{JMLR:v11:donmez10a}.}.
  
  In this work, we show how to combine neural networks for classification and $k$NN for regression is useful in reconstructing a new heavy Higgs boson decaying to $W^+W^-\to \ell^+\ell^{\prime -}+\nu_\ell\bar{\nu}_{\ell^\prime}$, $\ell(\ell^\prime)=e,\mu$, a fully leptonic final state with two escaping neutrinos, and its main SM backgrounds. We will show that the predicted mass of the charged leptons and neutrinos can be reliably used as a powerful new attribute to clean up the backgrounds further while enabling the selection of on-mass shell Higgs bosons.
 
 The work is organized as follows. In Section~\ref{sec:knn}, we describe the $k$NN regression algorithm; in Section~\ref{sec:reco}, we provide details of the combined construction of regressors and classifiers to identify the heavy Higgs boson and its main SM backgrounds, while in Section~\ref{sec:results} we present our final results in terms of improvement of the statistical significance of the signal hypothesis; Section~\ref{sec:conclusions} is devoted to conclusions and prospects.
 
\section{Details of the  $k$NN Regression}
\label{sec:knn}

 The $k$-nearest neighbors regressor~\cite{Altman} is a simple but effective algorithm for interpolation. First of all, we define a support dataset ${\cal S}=\{(\mathbf{X}_i,F(\mathbf{X}_i)),i=1,\cdots,N_s\}$, these are the exemplars which will be used to predict the value of the function of interest. Second, we define a distance metric, $Dist(\mathbf{X},\mathbf{Y})$, to decide which exemplars of  ${\cal S}$ are closer to a new point, $\mathbf{X}_{new}$, where $\mathbf{X}$, in our case, is a $\mathbb{R}^n$ vector. Third, we choose how many nearest neighbors to $\mathbf{X}_{new}$ will be used to compute $F(\mathbf{X}_{new})$, the target of our regression, according to a weighted mean
 \begin{equation}
     F(\mathbf{X}_{new}) = \frac{\sum_{m=1}^{k} F(\mathbf{X}_m)/Dist(\mathbf{X}_{new},\mathbf{X}_m)}{\sum_{m=1}^{k} 1/Dist(\mathbf{X}_{new},\mathbf{X}_m)}\; .
     \label{eq:kNN}
 \end{equation}
 Substituting  $Dist(\mathbf{X}_{new},\mathbf{X}_m)=1$ in the formula above corresponds to an arithmetic mean estimator for $F$. The weighted or arithmetic option will be decided in the tuning stage of the analysis.
 
 In principle, once we have chosen the distance metric, the number of nearest neighbors, $k$, used to compute $F(\mathbf{X}_{new})$ is the only hyperparameter of the algorithm. Note that this model has no parameters to be adjusted contrary to a neural network. This is the reason we do not need a training phase. However, the distance metric, $k$, and possibly other hyperparameters should be adjusted to get a good regressor by minimizing some error function. All $F(\mathbf{X}_m),\; m=1,\cdots,N_s$ are known thus, we are in the realm of supervised learning.
 
 In our case, the target function of the regression, $F$, is the leptonic $\ell^+\ell^{\prime -}\nu_\ell\bar{\nu}_{\ell^\prime}$, invariant mass, $M_{\ell\ell\nu\nu}$. The input of this function is the observable information obtained from the electrons and muons four-momenta, $p_e$ and $p_\mu$, respectively. The representation of the events was chosen as the energies and 3-momentum of the charged leptons plus high level functions construed from that low level information: $\mathbf{X}=(f_{ij}(p_\ell,p_{\bar{\ell}}),i=1,\cdots,N_{ev},\; j=1,\cdots,M)$ representing $N_{ev}$ events with $M$ features. 
 
  If the number of dimensions of the features space is large, distance-based models like $k$NN might perform poorly. For that reason, it is usual to project the features space onto a latent space of reduced dimensionality. There are various ways to do that. We chose to linearly transform the original features using a principal component analysis (PCA)~\cite{Deisenroth2020} and looking for the nearest neighbors in the transformed space of the first $P<M$ variables which best explain the variance of the data, $\mathbf{X}^{pca}={\cal T}_P(\mathbf{X})$.  We also adjust $P$ to obtain the best regressors. 
 
 One important ingredient of our method is based on the fact that experimental observations are organized in histograms of target variables. For a real-valued observable ${\cal O}$, what is truly compared against predictions are the number of events in pre-determined ranges of the observable, ${\cal O}_H = \{N_{ev,i} | {\cal O}\in [{\cal O}_i^{min}<{\cal O}<{\cal O}_i^{max}[\; ,i=1,\cdots,N_{bins}\}$. We found that predicting the bin where the event falls in histograms of $M_{\ell\ell\nu\nu}$ works better than predicting the value of $M_{\ell\ell\nu\nu}$ itself. We chose relatively large but fixed-sized bins. The binning itself, therefore, could be adjusted for performance mainly for large $M_{\ell\ell\nu\nu}$ where the number of events expected drops sharply. The regressor for the bins of the histogram of the $M_{\ell\ell\nu\nu}$ is given by
 \begin{equation}
     \hbox{bin of } M_{\ell\ell\nu\nu}({\cal T}_P(\mathbf{X}_{new})) = \frac{\sum_{m=1}^{k}\hbox{bin of }M_{\ell\ell\nu\nu}({\cal T}_P(\mathbf{X}_m))/Dist({\cal T}_P(\mathbf{X}_{new}),{\cal T}_P(\mathbf{X}_m)) }{\sum_{m=1}^{k} 1/Dist({\cal T}_P(\mathbf{X}_{new}),{\cal T}_P(\mathbf{X}_m))}\; .
     \label{eq:kNN_binned}
 \end{equation}

Let us now construct the regressors for the signal and the backgrounds. 

\section{Reconstruction of fully leptonic resonances}
\label{sec:reco}

The dataset consists of 400000 simulated signal events $pp\to H_2\to W^+W^-\to \ell^+\ell^{\prime -}+\nu_\ell\bar{\nu}_{\ell^\prime}$, $\ell(\ell^\prime)=e,\mu$, where $H_2$ is a new Higgs boson produced via gluon fusion, for each one of the three different mass values: 1, 1.5 and 2 TeV and two fixed total $H_2$ width, 1\% and 10\% of the mass parameter, totaling 2.4 million signal events. The dataset also contains 5.2 million of the corresponding SM backgrounds evenly split into four processes, as we discuss ahead. Our goal is to show that the resonance can be reliably reconstructed. Using it can boost both ML classifiers' accuracy and other metrics and the signal significance compared to a baseline classifier without the $\mllvv$ regression. The true value of the statistical significance is actually of minor importance to us, so we fix the number of signal events to illustrate our method. Our sole supposition is that the leptons plus neutrinos signals are dominated by the $WW$ mode with negligible interference with the corresponding SM backgrounds.\footnote{A non-negligible interference with the SM Higgs boson is expected with wide scalar resonances of masses below 1 TeV or so. This should not pose any difficulties for the $k$NN regression, however.}.
 
 We consider the following background sources in our analysis: (1) the dominant irreducible component, $pp\to W^+W^-$, (2) the subdominant irreducible, $pp\to ZZ(\gamma^*)$, (3) the dominant reducible contribution, $pp\to t\bar{t}\to W^+W^- b\bar{b}$. 
 All the signals and backgrounds partonic events are simulated at leading order using \texttt{MadGraph5}~\cite{Alwall:2014hca}. Hadronization is simulated with \texttt{Pythia8}~\cite{Sjostrand:2007gs}, while detectors effects are simulated with \texttt{Delphes3}~\cite{deFavereau:2013fsa}. 
 
 The partonic events are used to obtain the ground truth $\mllvv$ distributions once the neutrinos momenta are available. Note that this distribution explicitly assumes that missing energy is all due to escaping neutrinos produced in the hard scattering, but not the misreconstruction of observable momenta or the missing of other particles. However, the leptons momenta and the event's missing energy, which feed the algorithms, include all the simulated effects. This is another reason to construct a regressor for the distribution bins. For sufficiently large bins, the mismatch between the partonic $\mllvv$ and $M_{\ell\ell\; \met}$ can be more easily accommodated without affecting the quality of the regression.
 
 \begin{figure}[t!]
     \centering
     \includegraphics[scale=0.34]{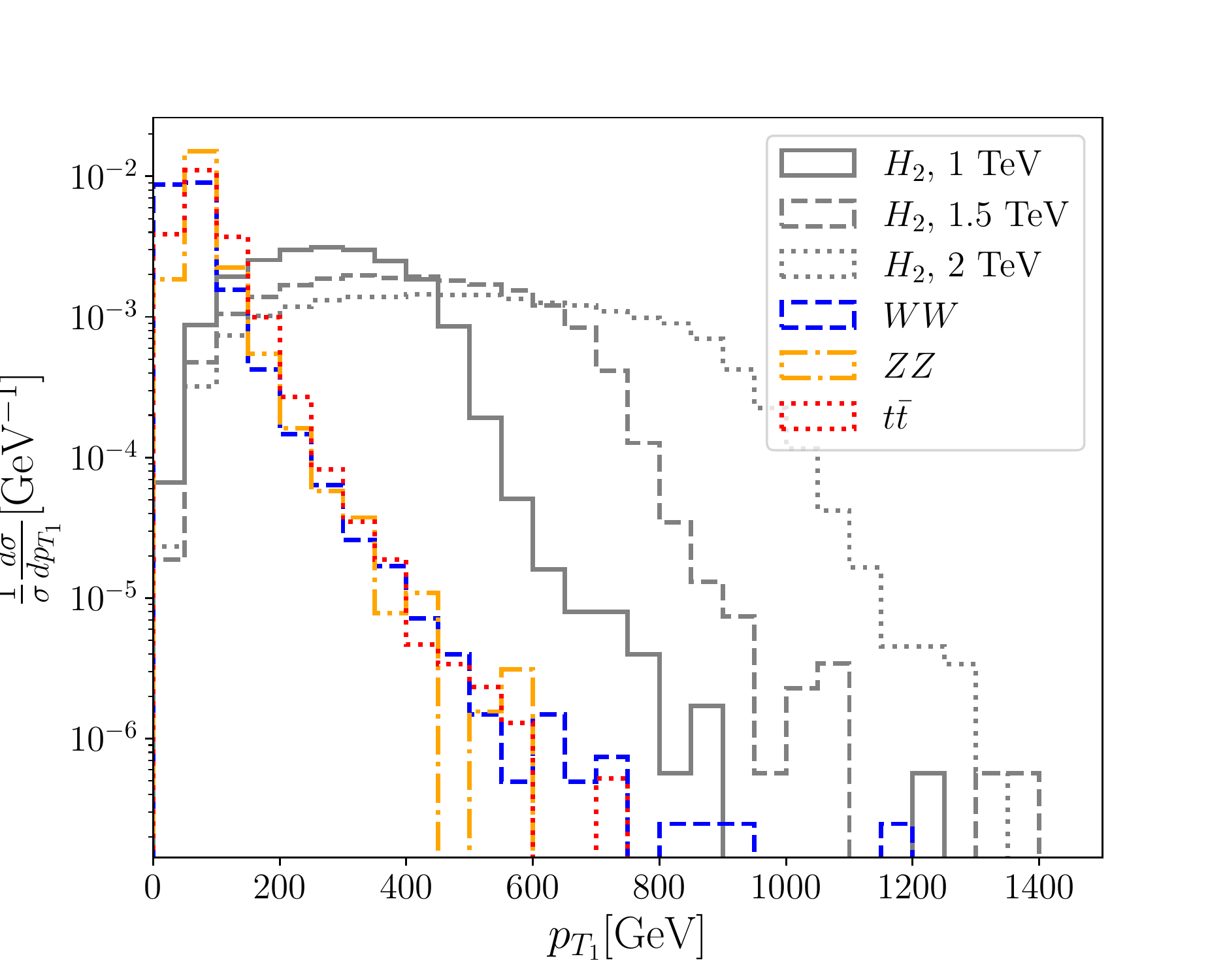}
     \includegraphics[scale=0.34]{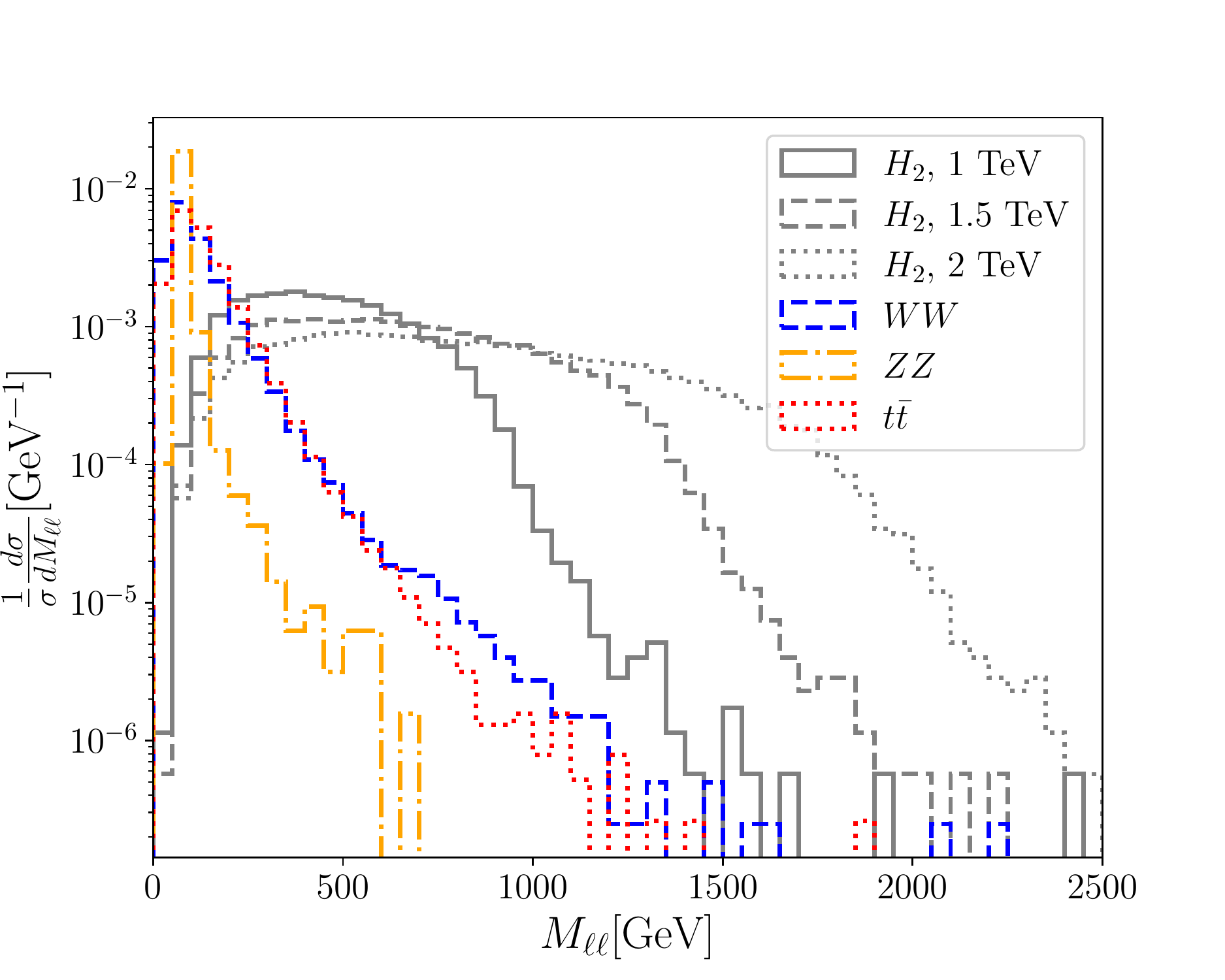}\\
     \includegraphics[scale=0.34]{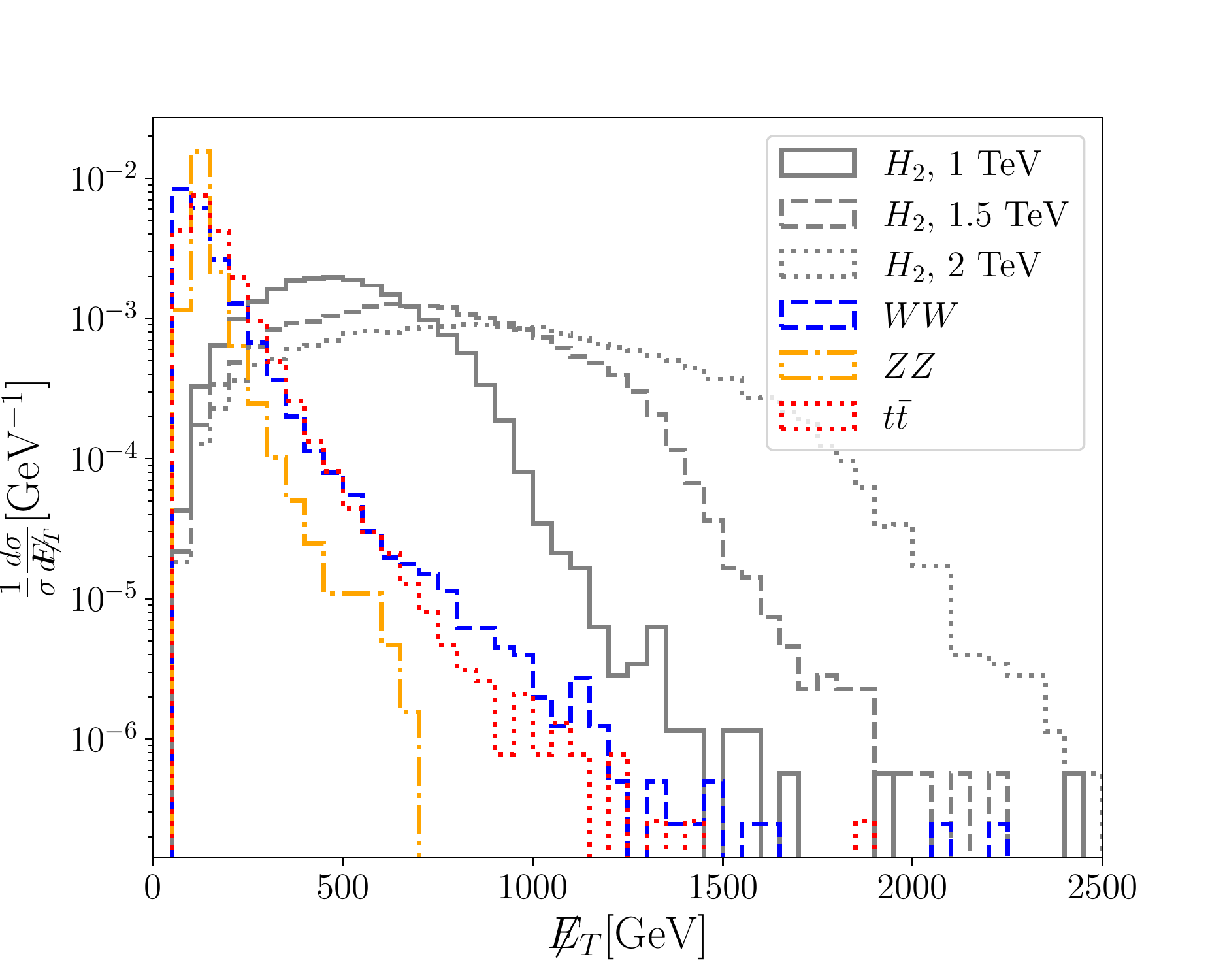}
     \includegraphics[scale=0.34]{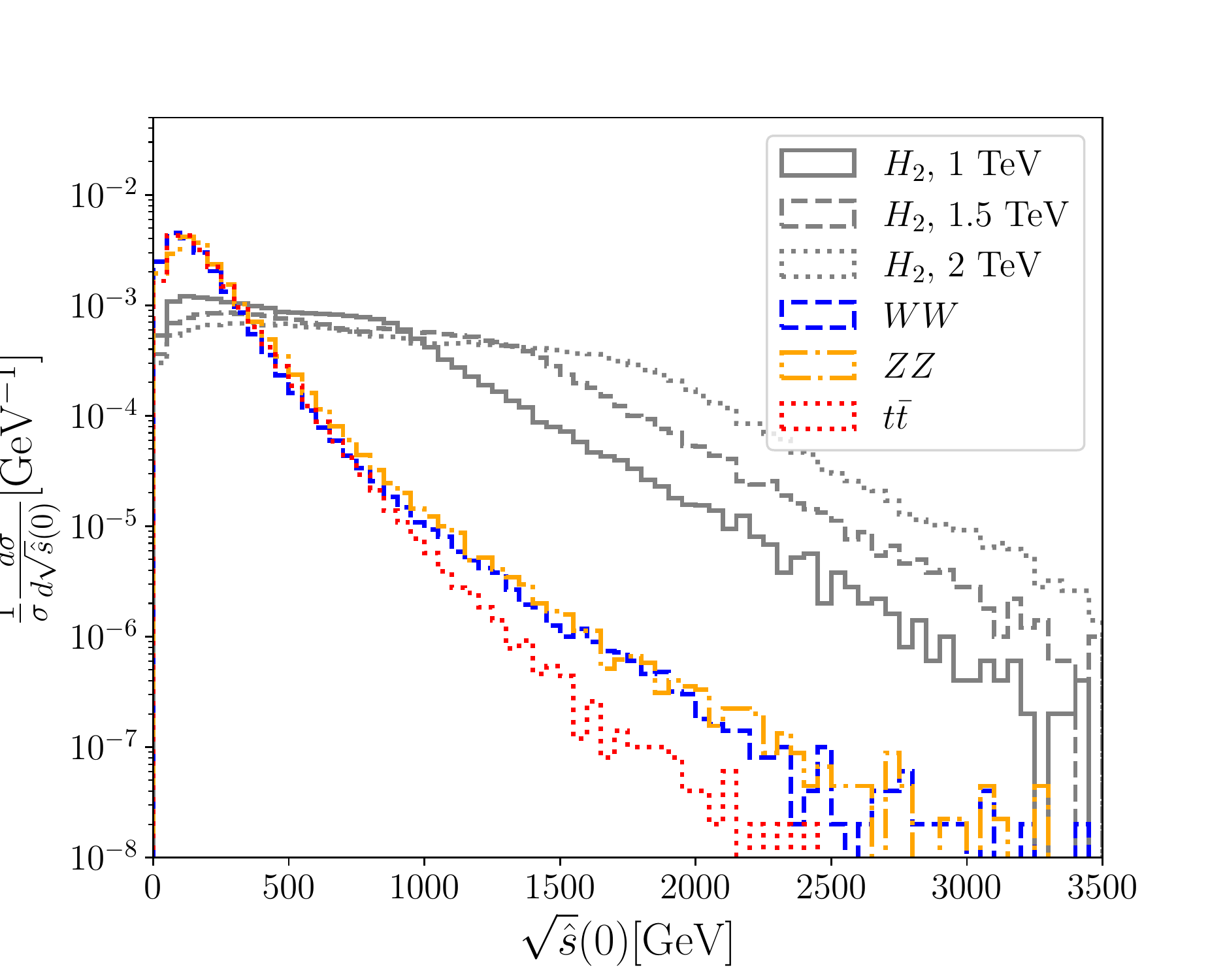}\\
     \includegraphics[scale=0.34]{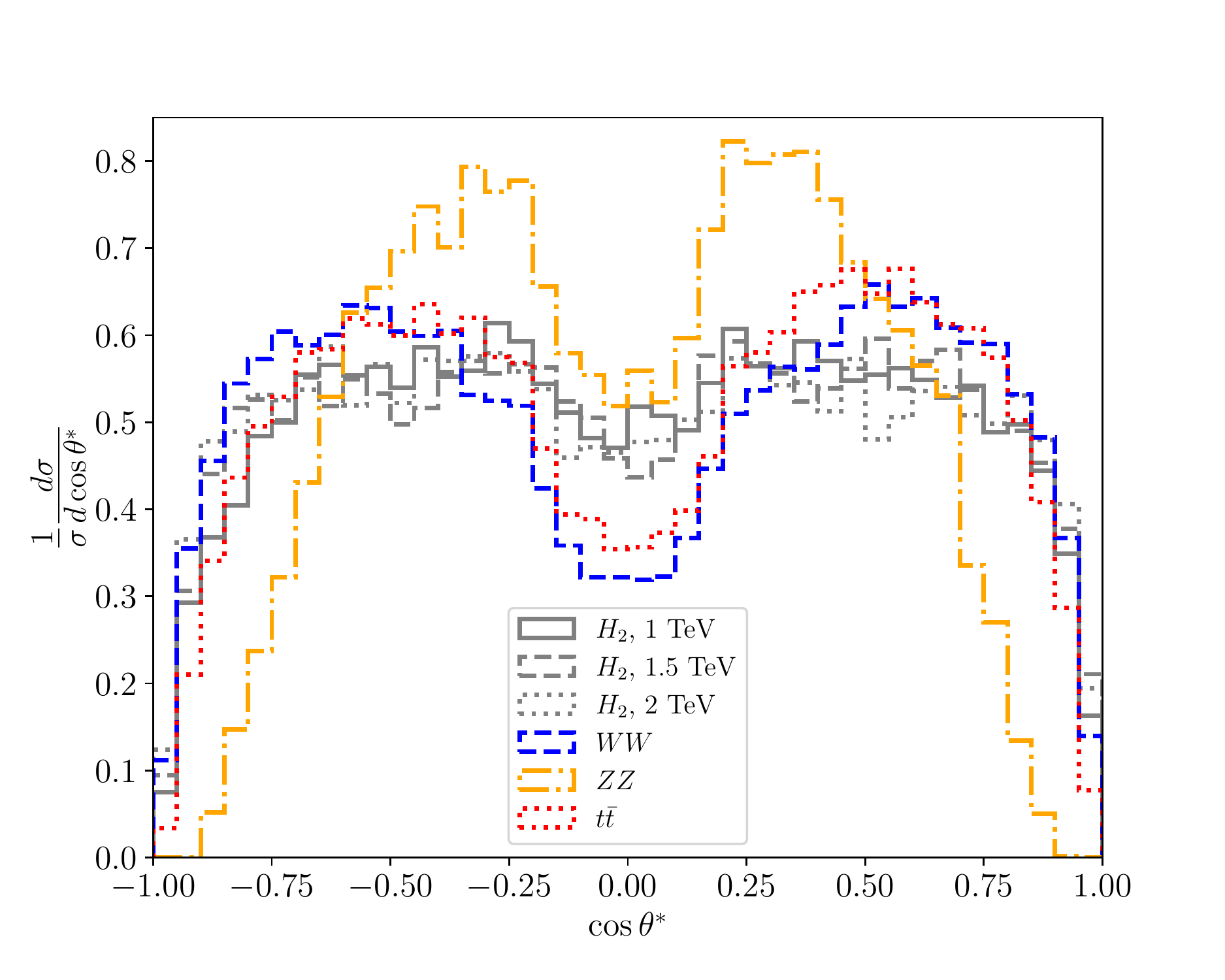}
     \includegraphics[scale=0.34]{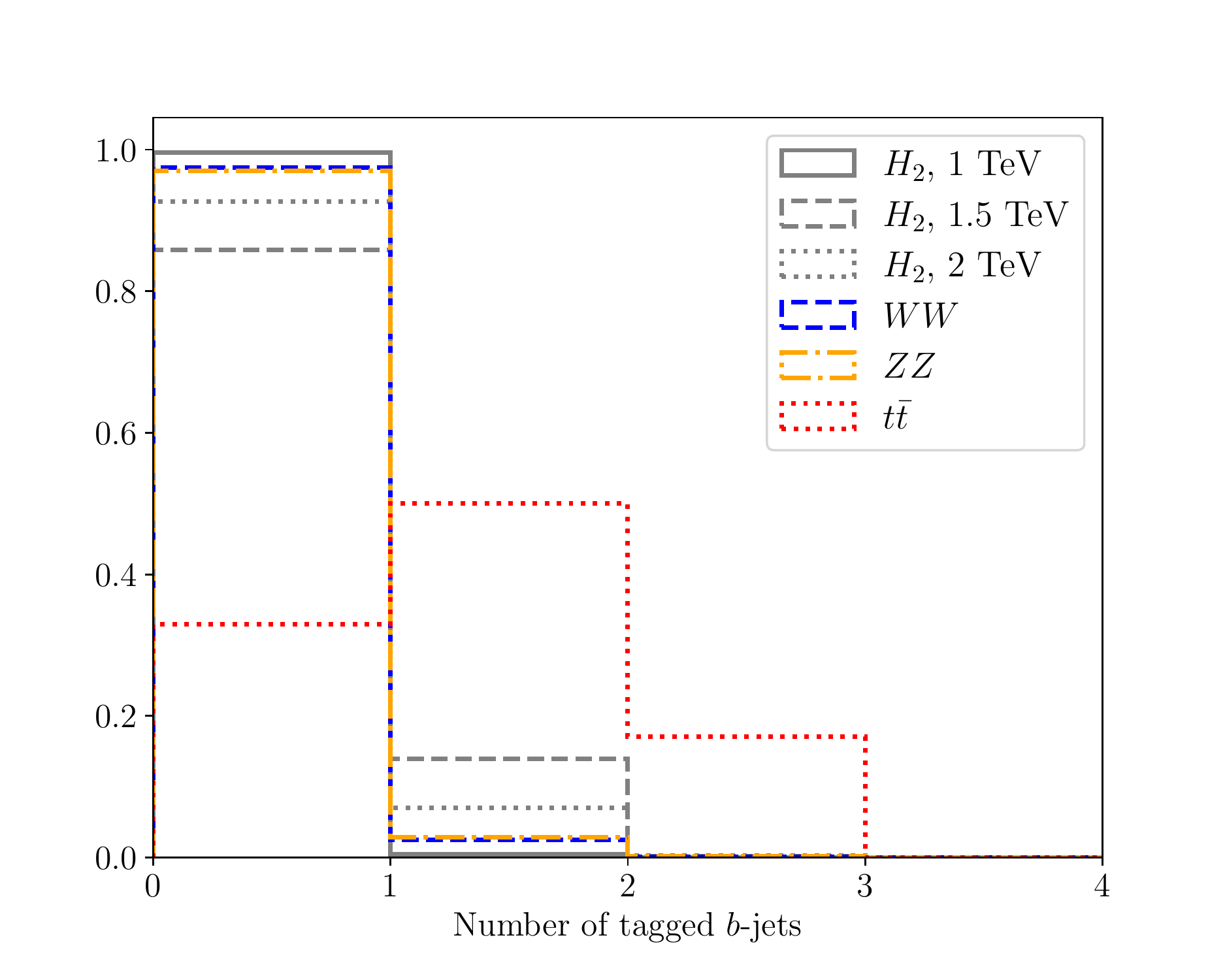}\\
     \includegraphics[scale=0.34]{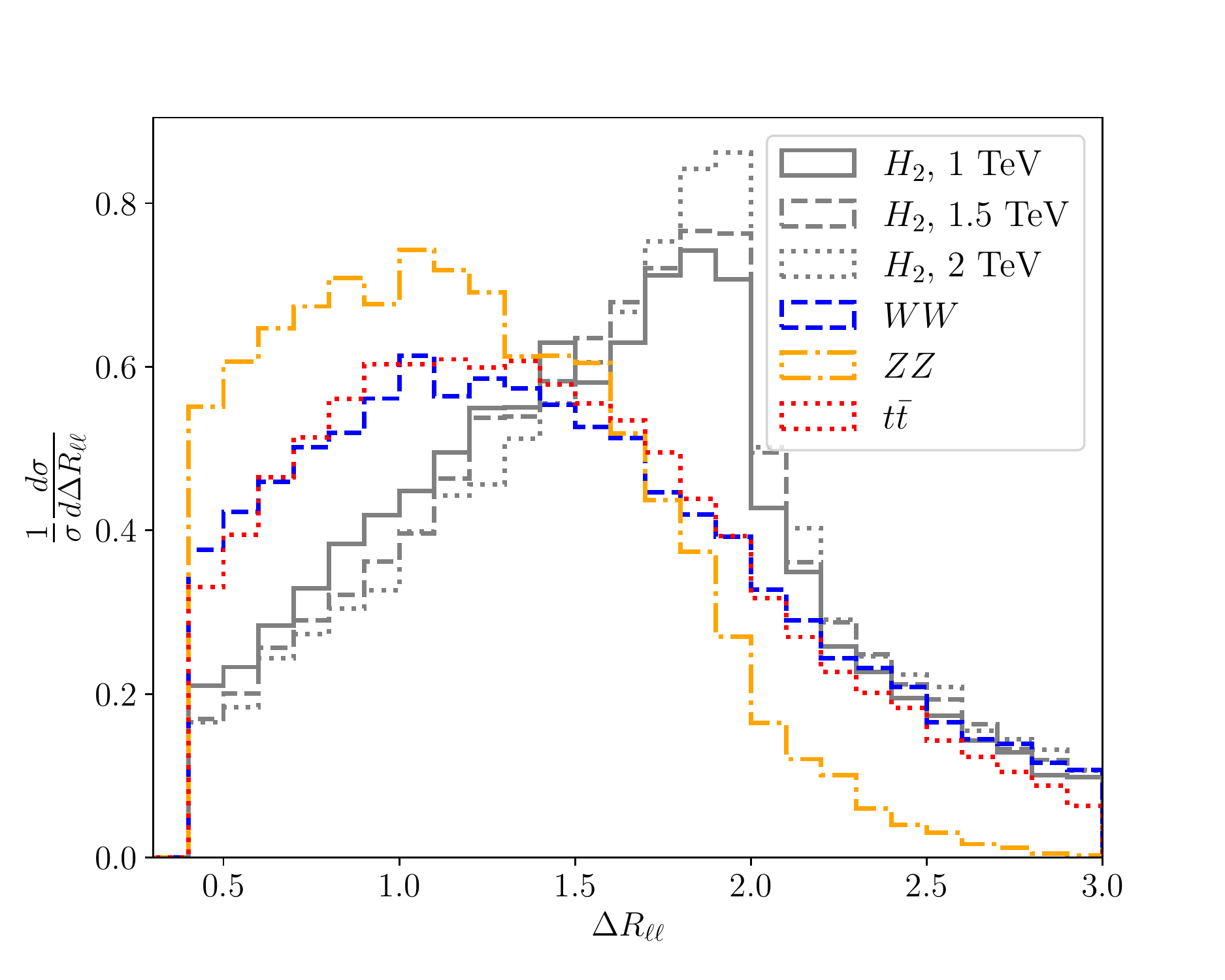}
     \includegraphics[scale=0.34]{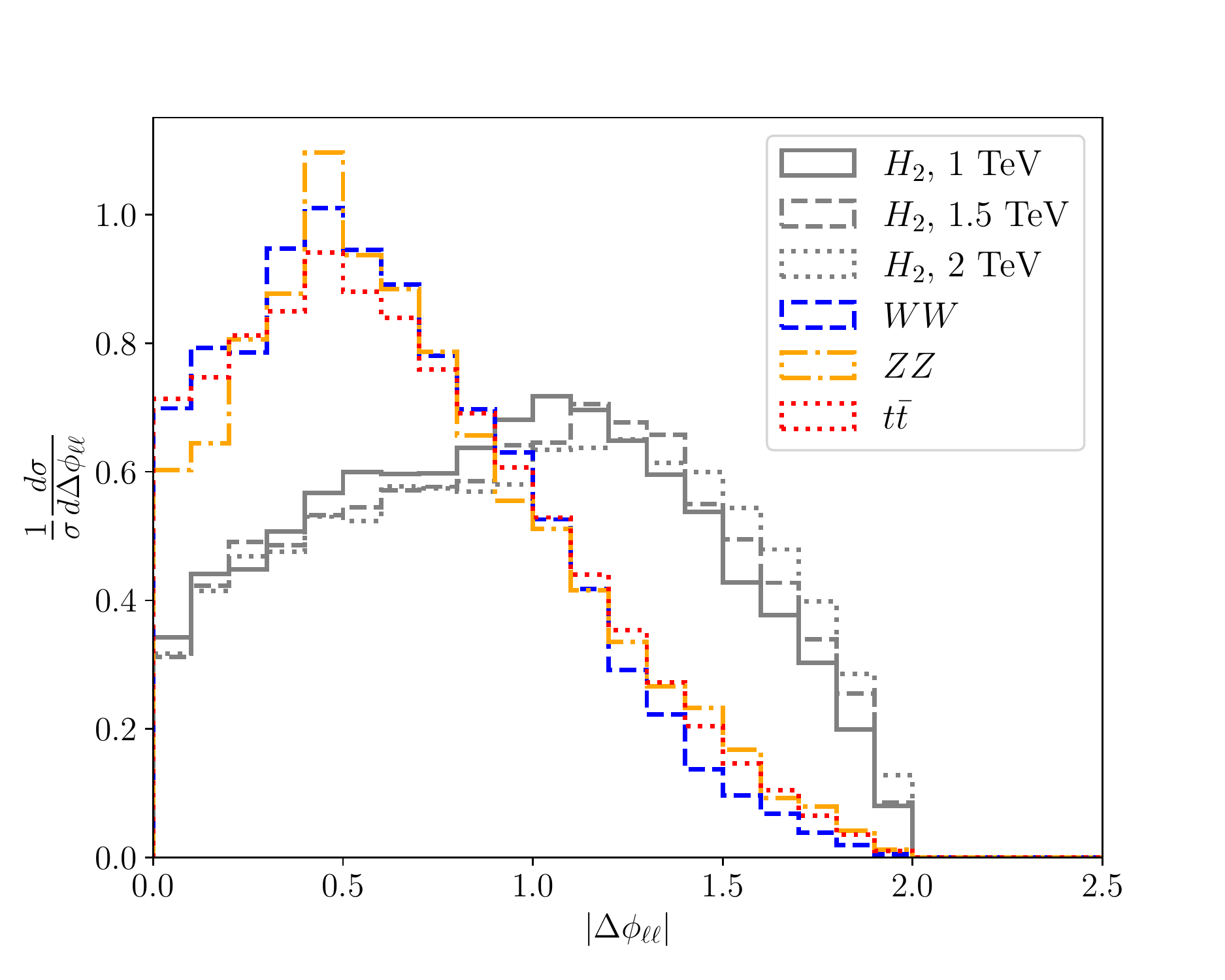}
     \caption{Some of the kinematic distributions of Higgs bosons and its corresponding SM backgrounds chosen to represent the events for regression and classification.}
     \label{fig:distr}
 \end{figure}
 We adopt the following basic acceptance cuts to select events with two opposite charge leptons and missing energy 
 \begin{equation}
  p_{T,\ell} > 20\; \hbox{ GeV},\;\; |\eta_\ell| < 2.4,\;\; \Delta R_{\ell\ell}>0.4,\;\; M_{\ell\ell} > 30\hbox{ GeV},\;\; \met > 40\hbox{ GeV},\;\; |\Delta\eta_{\ell\ell}|<3.0,
  \label{eq:cuts}
 \end{equation}
  where $p_{T,\ell}$ and $\eta_\ell$ denote the transverse momentum and pseudo-rapidity of the leptons, respectively, while $M_{\ell\ell}$, $\met$ and $\Delta R_{\ell\ell}$ denote the invariant mass of the charged leptons, the missing transverse energy and the distance in the $\eta\times\phi$ plane of the event. The last cut, on the rapidity gap between the charged leptons, was imposed to suppress weak boson fusion backgrounds, which are neglected in the subsequent analysis. The $M_{\ell\ell}$ helps to suppress the low mass leptons backgrounds from $Z\gamma^*$, which showed to be a source of contamination among the events classes.

 
 In Fig.~\eqref{fig:distr}, we show the distributions of some features chosen to represent the events and predict their classes and $\mllvv$. Along with the energies and the components of the 3-momenta of the charged leptons, we also include their transverse momentum, and the following variables:
 \begin{itemize}
     \item $M_{\ell\ell}$, the charged leptons invariant mass,
     \item $\met$, the missing transverse energy,
     \item $\Delta R_{\ell\ell}=\sqrt{(\Delta\eta_{\ell\ell})^2+(\Delta\phi_{\ell\ell})^2}$, where $\Delta\eta_{\ell\ell}$ and $\Delta\phi_{\ell\ell}$ represent the rapidity and azimuthal angle differences between the charged leptons,
     \item $\cos\theta^* = \tanh\left(\frac{\Delta\eta_{\ell\ell}}{2}\right)$, proposed in Ref.~\cite{Barr:2005dz},
     \item $\sqrt{\hat{s}}(0)=\sqrt{E_{\ell\ell}^2-p_{T,\ell\ell}^2} + \not\!\! E_T$, proposed in Ref.~\cite{Konar:2010ma}, 
     \item the number of jets tagged as a bottom jet to suppress $t\bar{t}$ events.
 \end{itemize}
 
 For each class, we construct a regressor function according to Eq.~\eqref{eq:kNN_binned}. 
 At this stage, we employed 0.9 and 1.2 million events for signals and backgrounds, respectively. To ensure that the dataset's size would not play a role in the results, we separated 80\% of that data for tuning the regressors. We adjusted, with a grid search, the number of nearest neighbors, $k$, the distance metric\footnote{For a good account on $k$NN and its options, including the distance metrics available, see
   \href{https://scikit-learn.org/stable/modules/neighbors.html}{sklearn page}.
   Our results are, in fact, insensitive to the distance metric option as explained in the text, we show them just for completeness.}, $Dist$, the number of PCA transformed variables, $P$, and the weighted or arithmetic option in Eq.~\eqref{eq:kNN_binned} in order to minimize the mean square error between the predicted and the true binned $\mllvv$ distributions. The space of hyperparameters in the grid search is the following
 \begin{eqnarray}
&& k\in [1,5],\;\; Dist\in \{\hbox{\texttt{Minkowsky,Manhattan,Chebyshev}}\},\nonumber\\
&& P\in [1,8],\;\; \hbox{weight}\in  \{\hbox{\texttt{uniform},\texttt{weighted}}\}.
\label{eq:tuning}
 \end{eqnarray}
 
\begin{figure}[t!]
     \centering
     \includegraphics[scale=0.3]{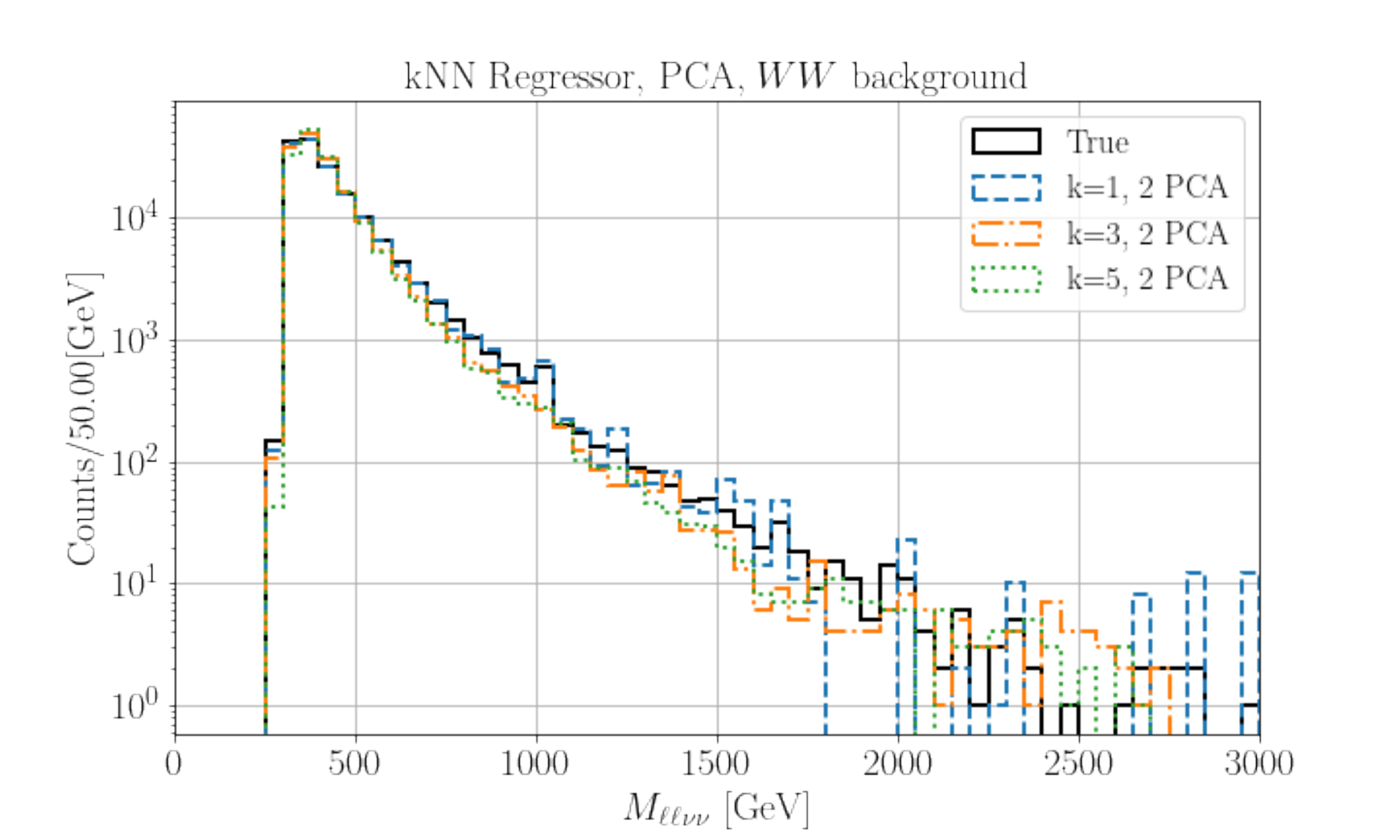}
     \includegraphics[scale=0.3]{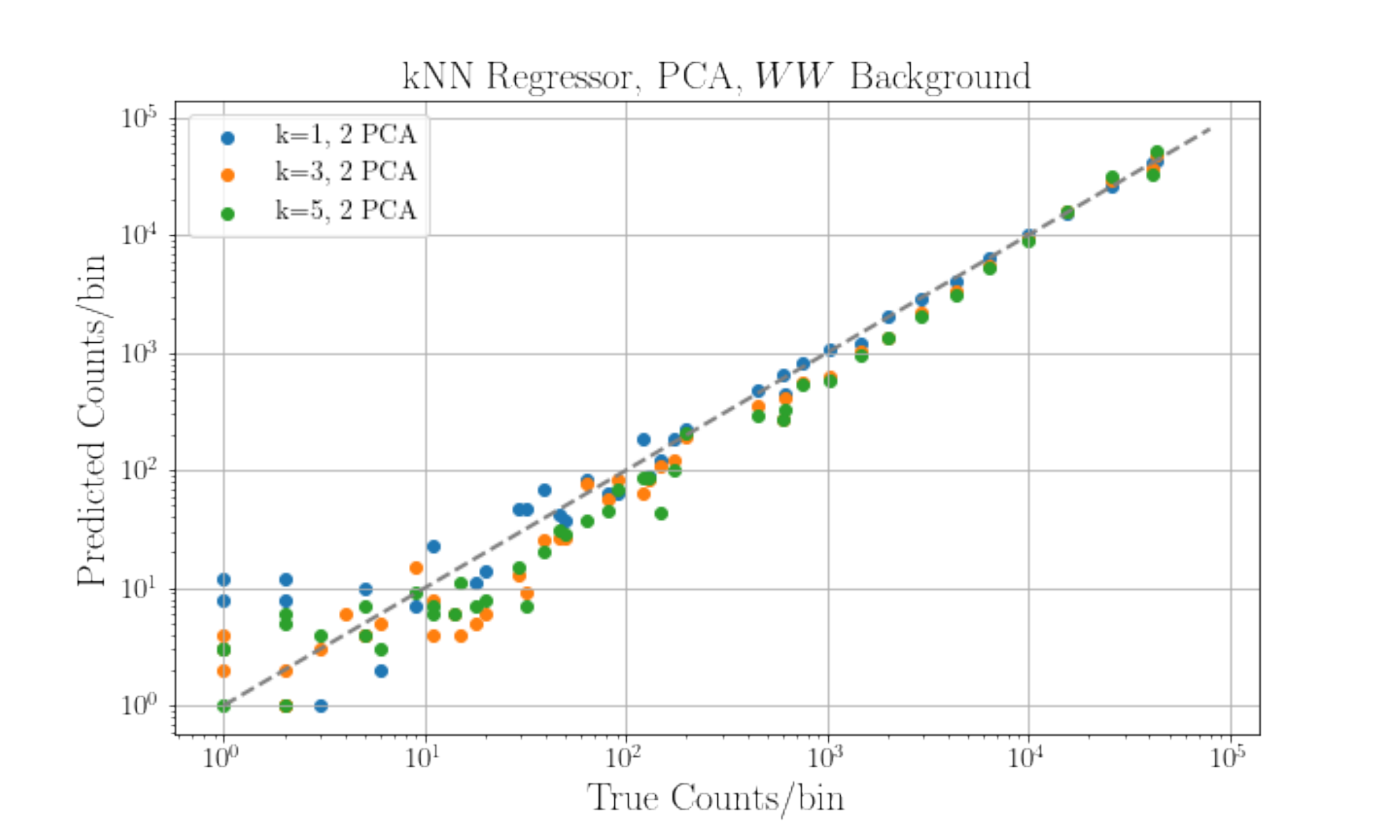}\\
     \includegraphics[scale=0.3]{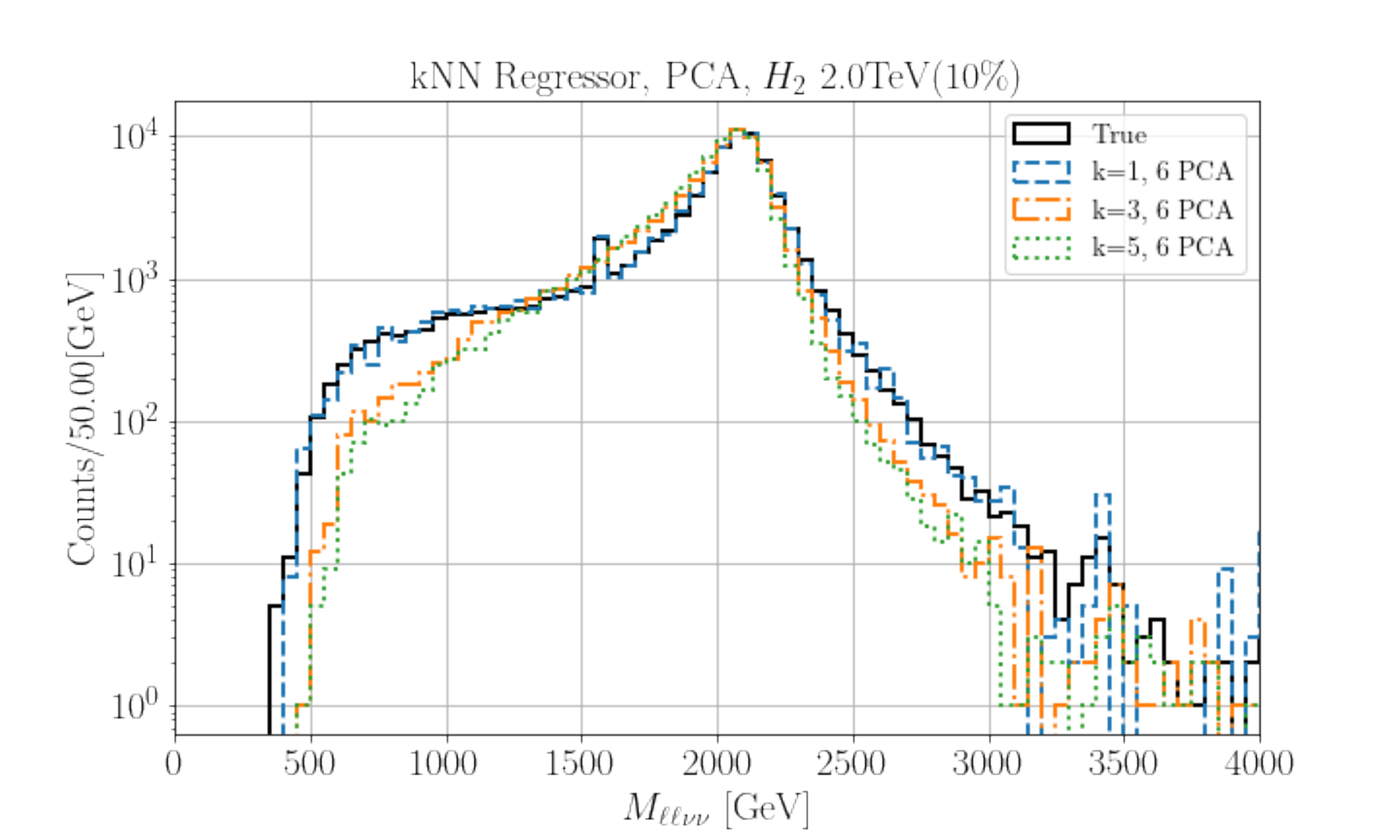}
     \includegraphics[scale=0.3]{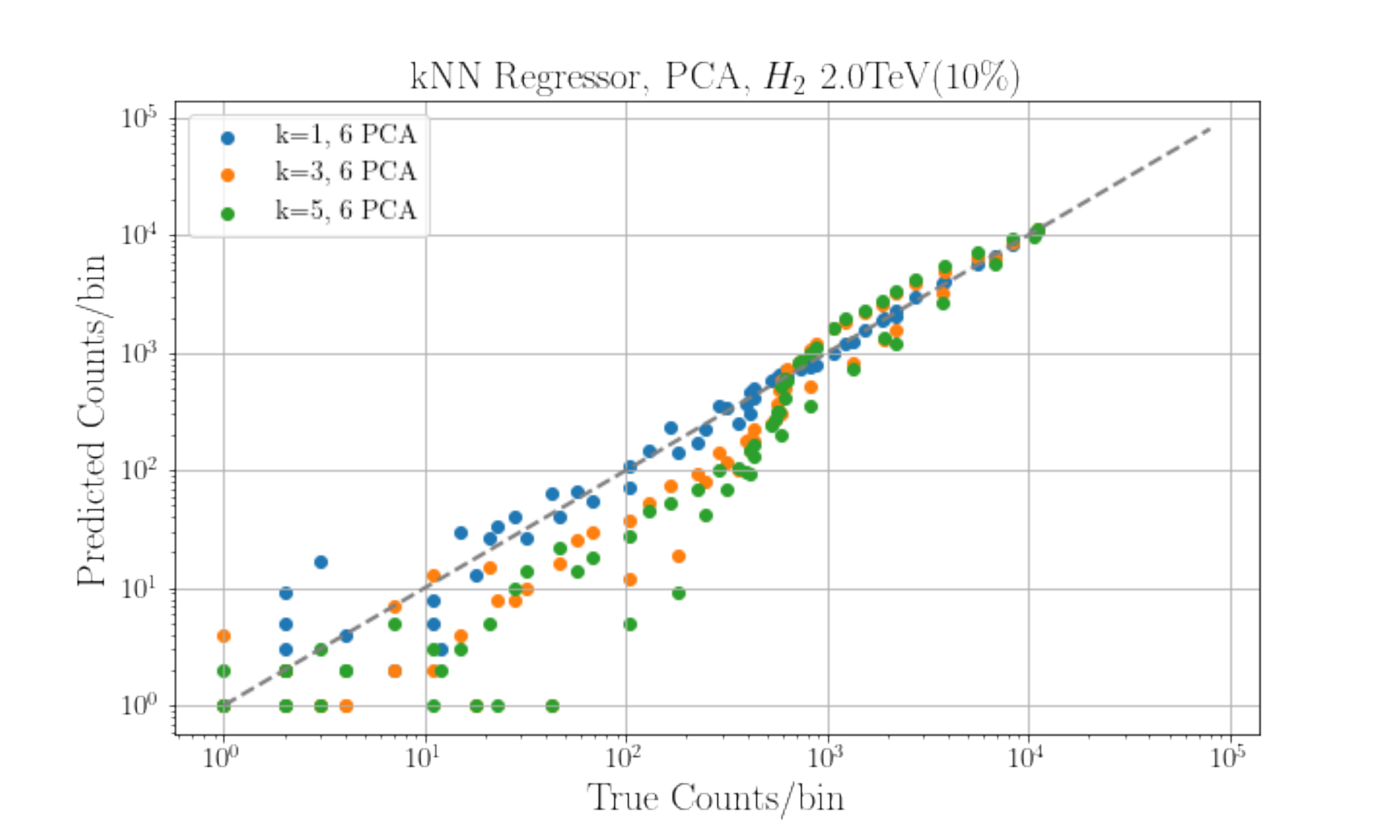}\\
     \includegraphics[scale=0.3]{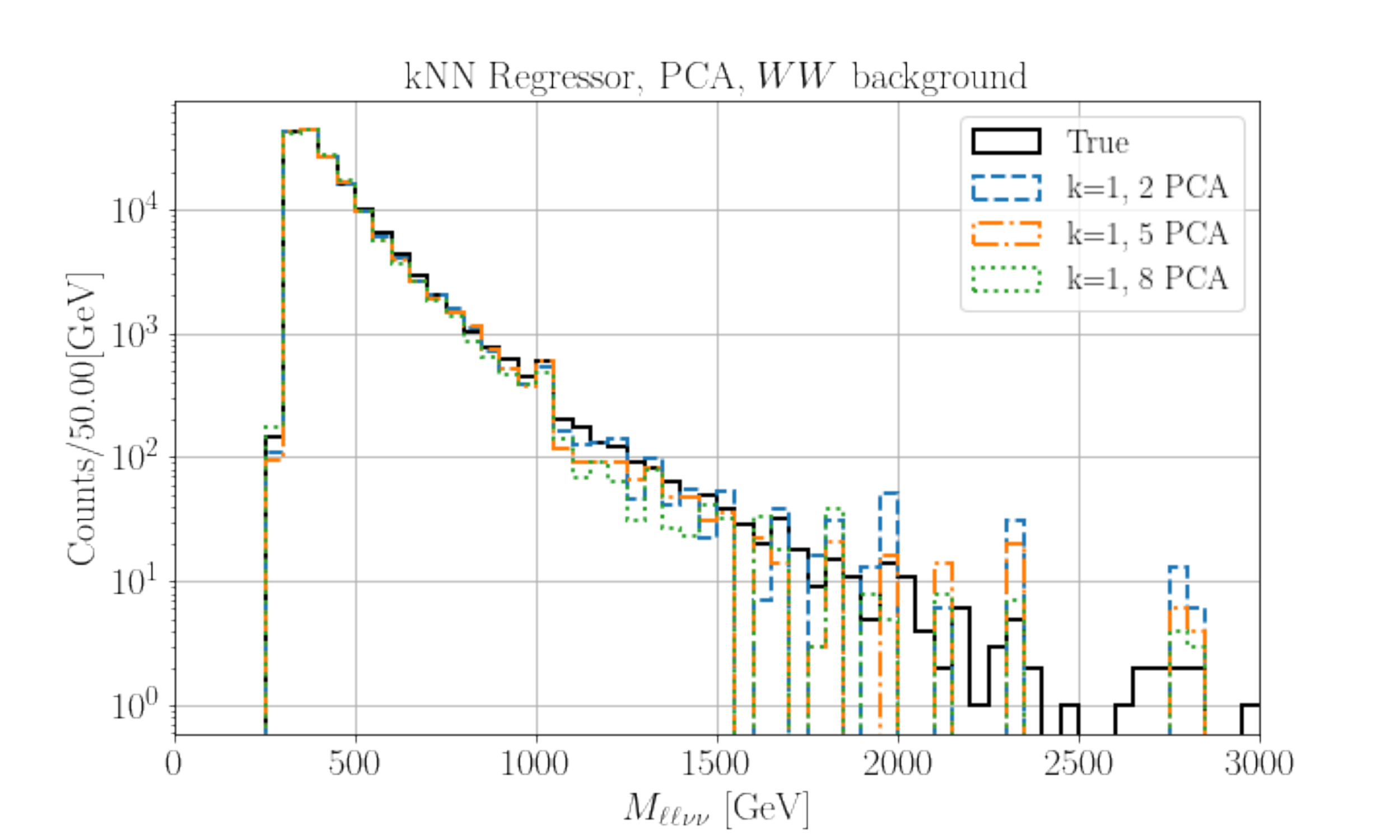}
     \includegraphics[scale=0.3]{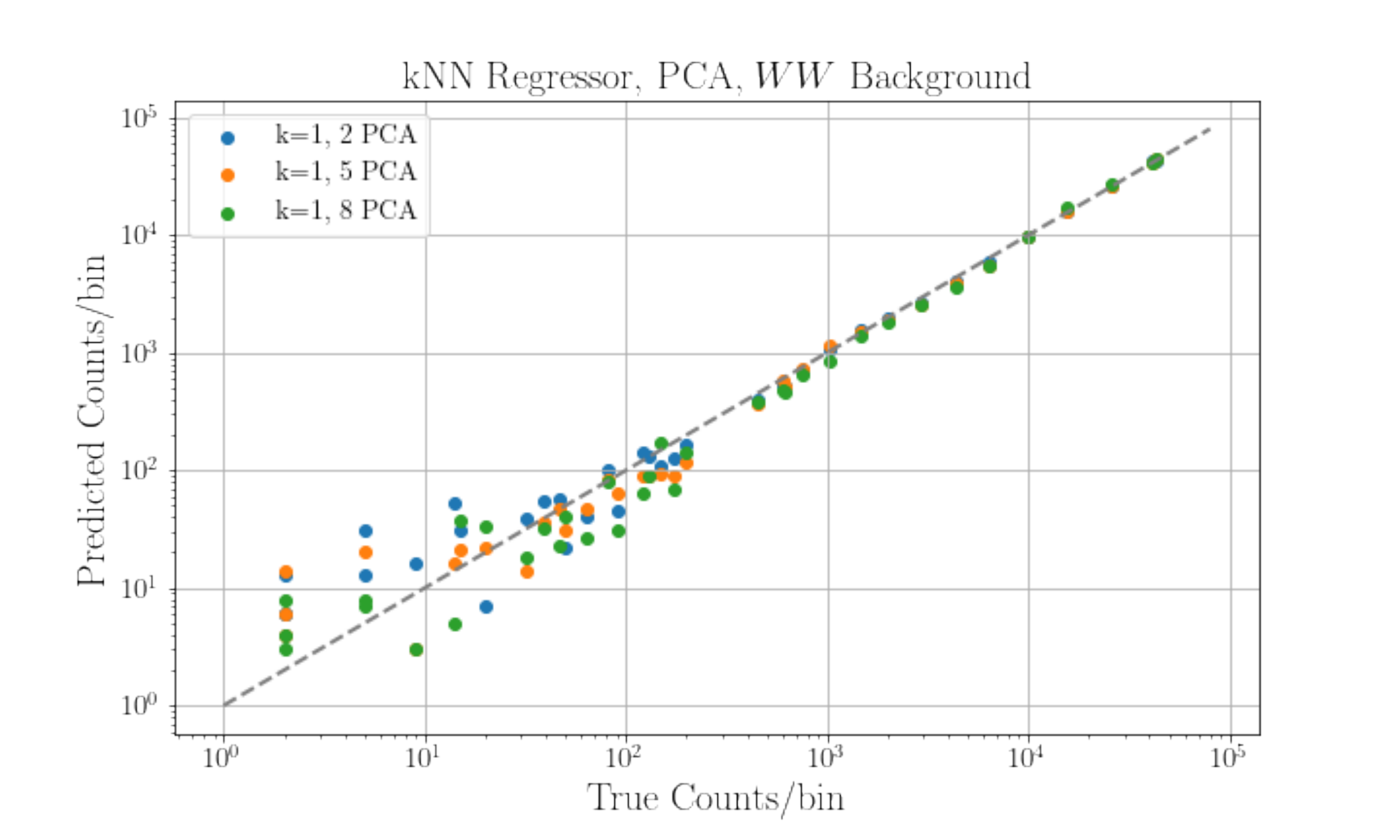}\\
     \includegraphics[scale=0.3]{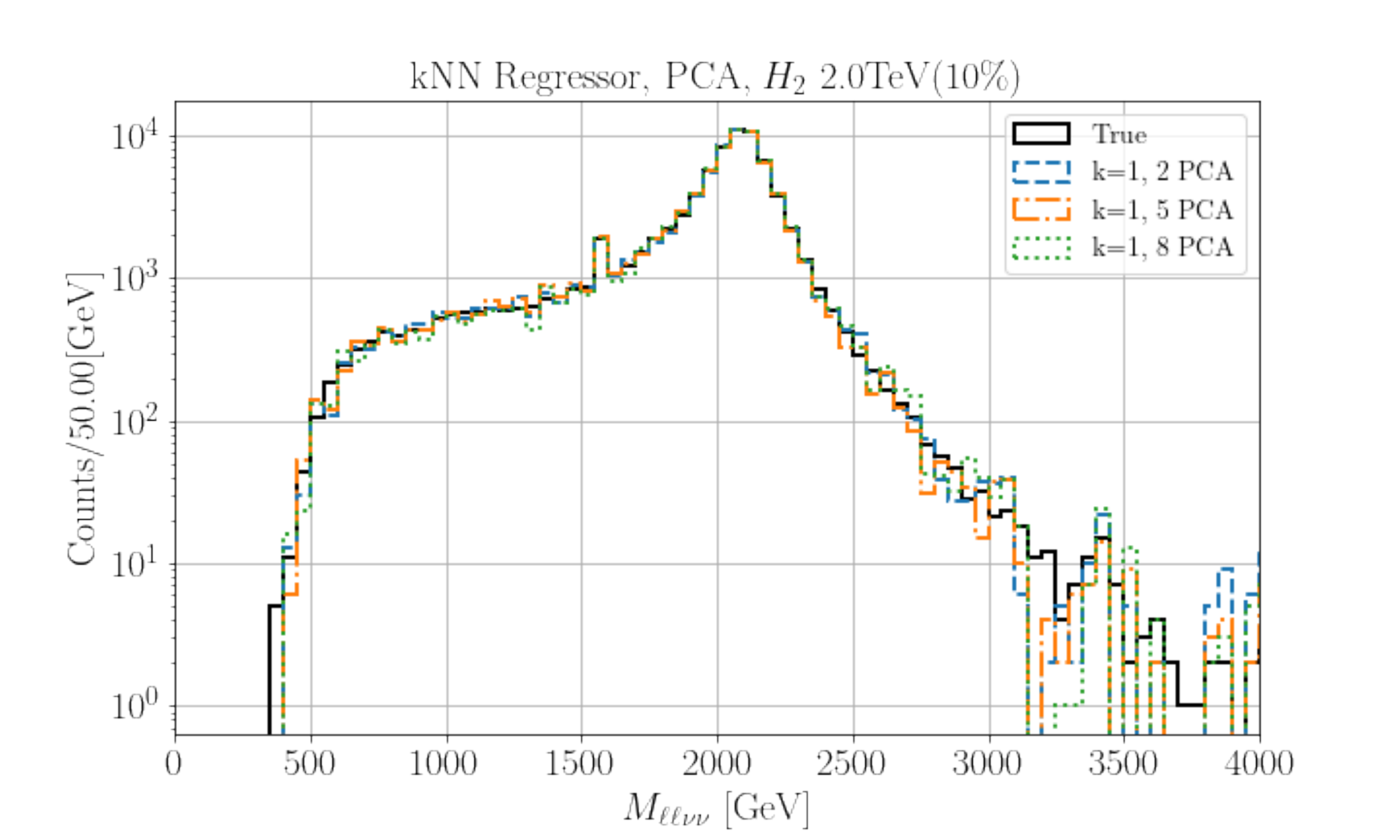}
     \includegraphics[scale=0.3]{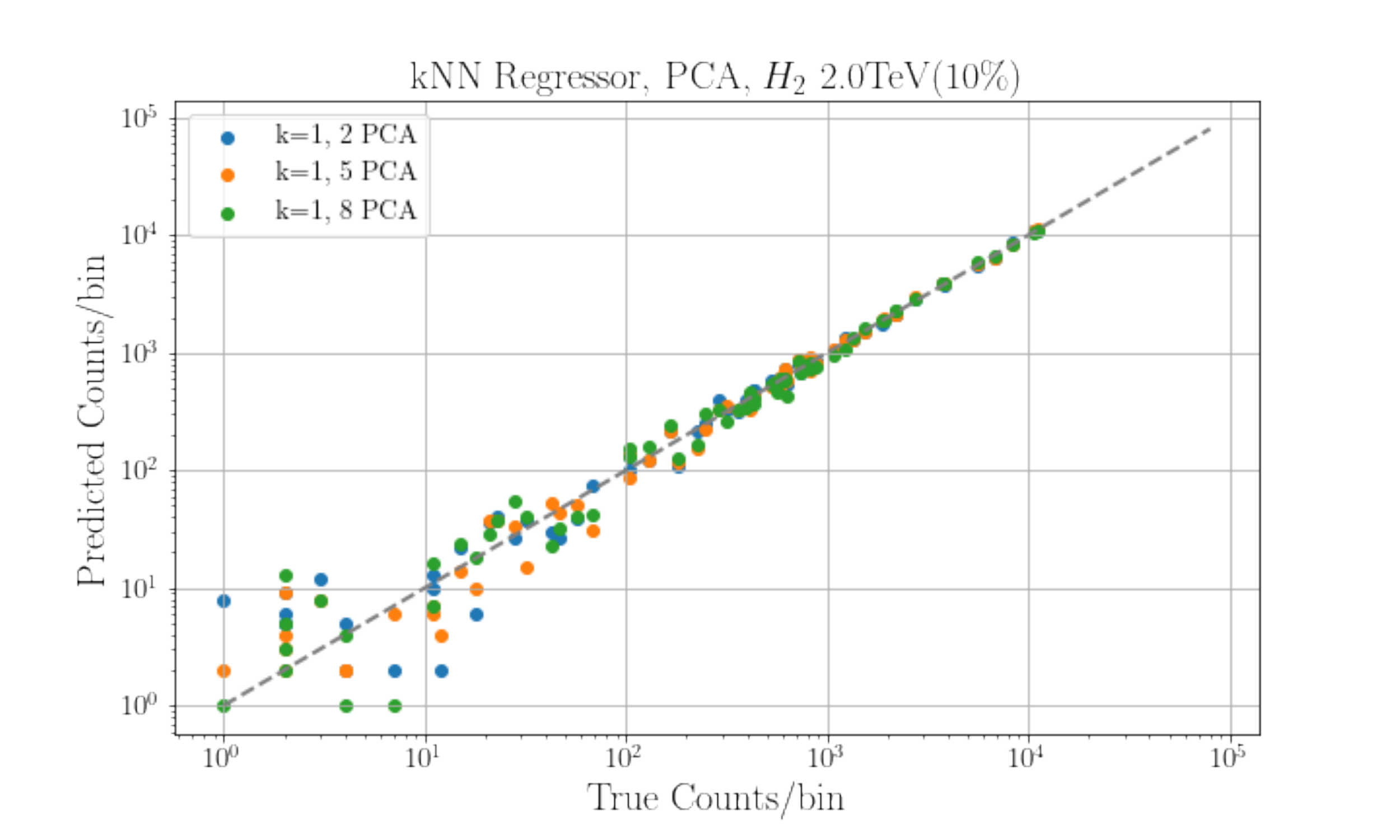}
     
     \caption{Results for the tuning of the number of nearest neighbors, $k$, and number of principal components (PCA) of the latent space. In the four upper panels we display $k=$1, 3, and 5, keeping PCA fixed at its best value. In the four lower panels we display PCA=2, 5, and 8, keeping $k$ fixed at its best value. }
     \label{fig:tuning}
 \end{figure}

 We display, in Fig.~\eqref{fig:tuning}, some results of the tuning of the number of nearest neighbors, $k$, and the number of principal components to demonstrate the quality of the $k$NN regression for the cases of the SM $WW$ background and a 2 TeV Higgs boson. The other backgrounds and signals present very similar behavior. The best hyperparameters were chosen as those with the smaller mean squared error (MSE) between the true and predicted histograms of the target variable.

 The $k$NN regressor is robust against most parameter variations while being very accurate for predictions. Overall, for all backgrounds and the signals, the nearest neighbor to a new point in the latent space of PCA transformation is the most accurate prediction for our target variable. We tested various alternatives to $k$NN as gradient boosting and neural networks regressors, and the nearest neighbors approach showed itself superior in approximating the true distribution of masses. We also found that neural networks present an improved generalization performance across classes compared to other algorithms, especially the $k$NN algorithm, which is very dependent on the class of the event. For example, we found that training a neural network with $WW$ backgrounds might be useful to obtain $\mllvv$ for the other classes, especially the backgrounds, but its performance on signal events is still not competitive with much simpler proxy variables that correlate with the resonance mass, as $\sqrt{\hat{s}}(0)$~\cite{Konar:2010ma} and other transverse mass variables. The significant advantage of algorithms with good generalization performance is being agnostic towards the other classes, depending less on the previous knowledge of the types of events.
  \begin{figure}[t!]
     \centering
     \includegraphics[scale=0.4]{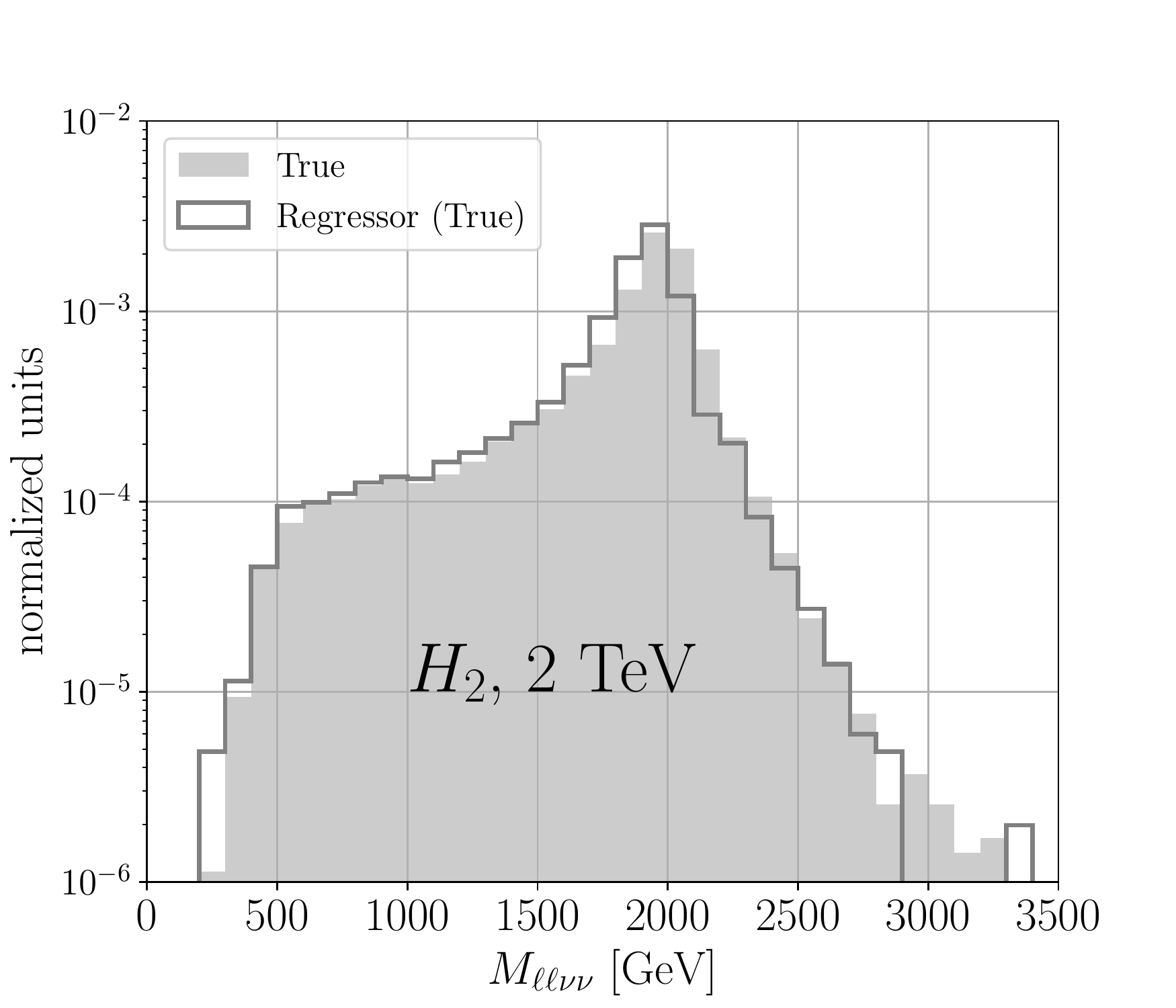}
     \includegraphics[scale=0.4]{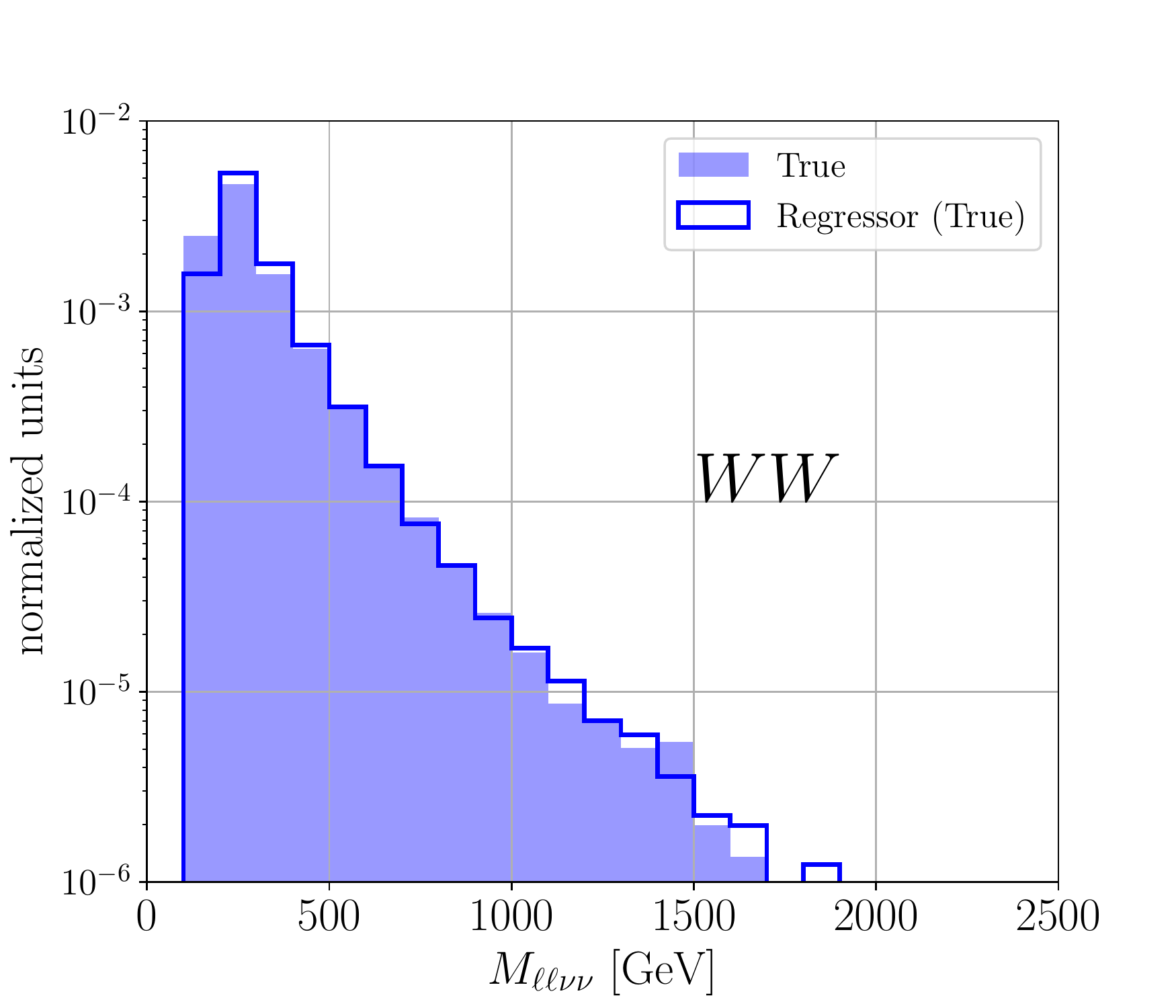}\\
     \includegraphics[scale=0.4]{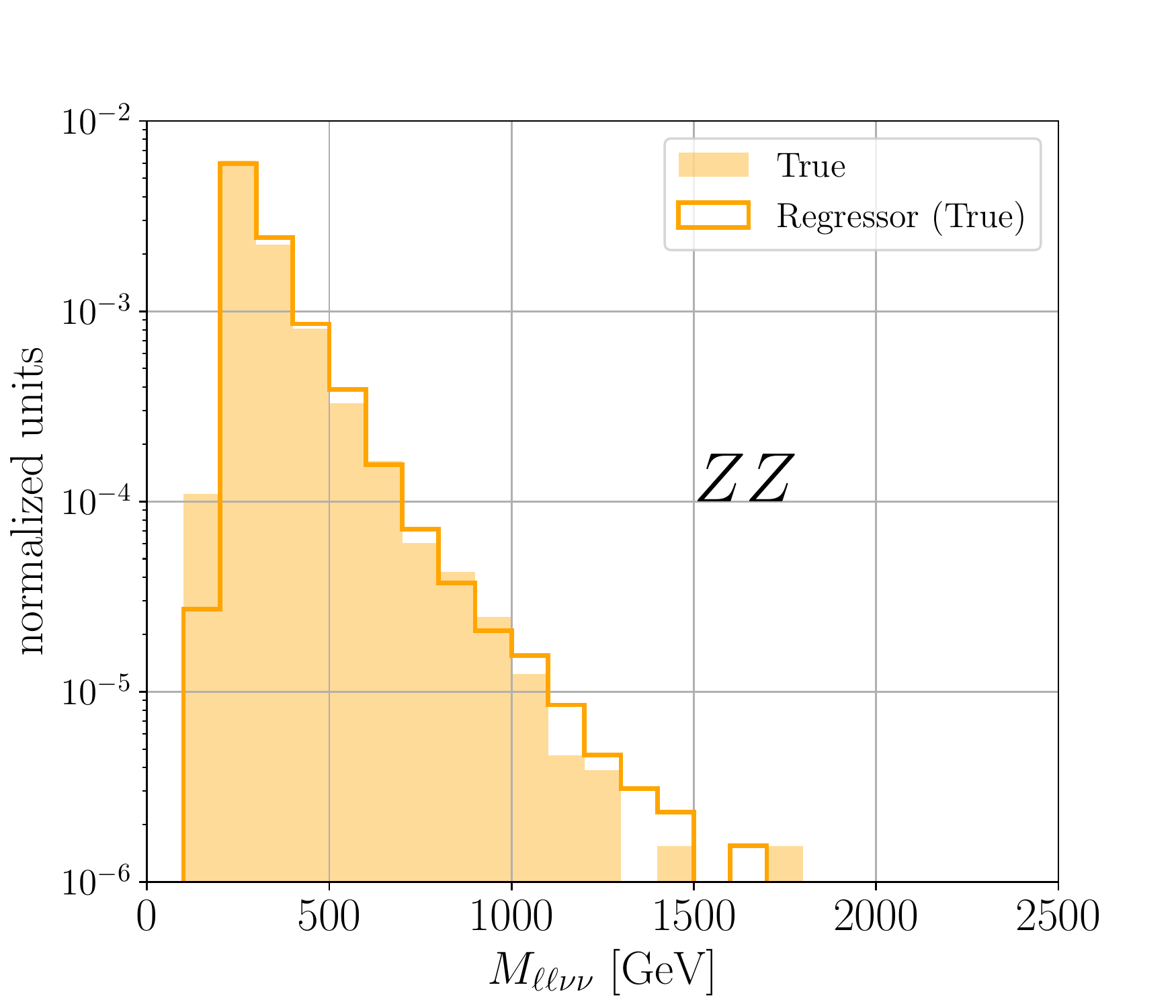}
     \includegraphics[scale=0.4]{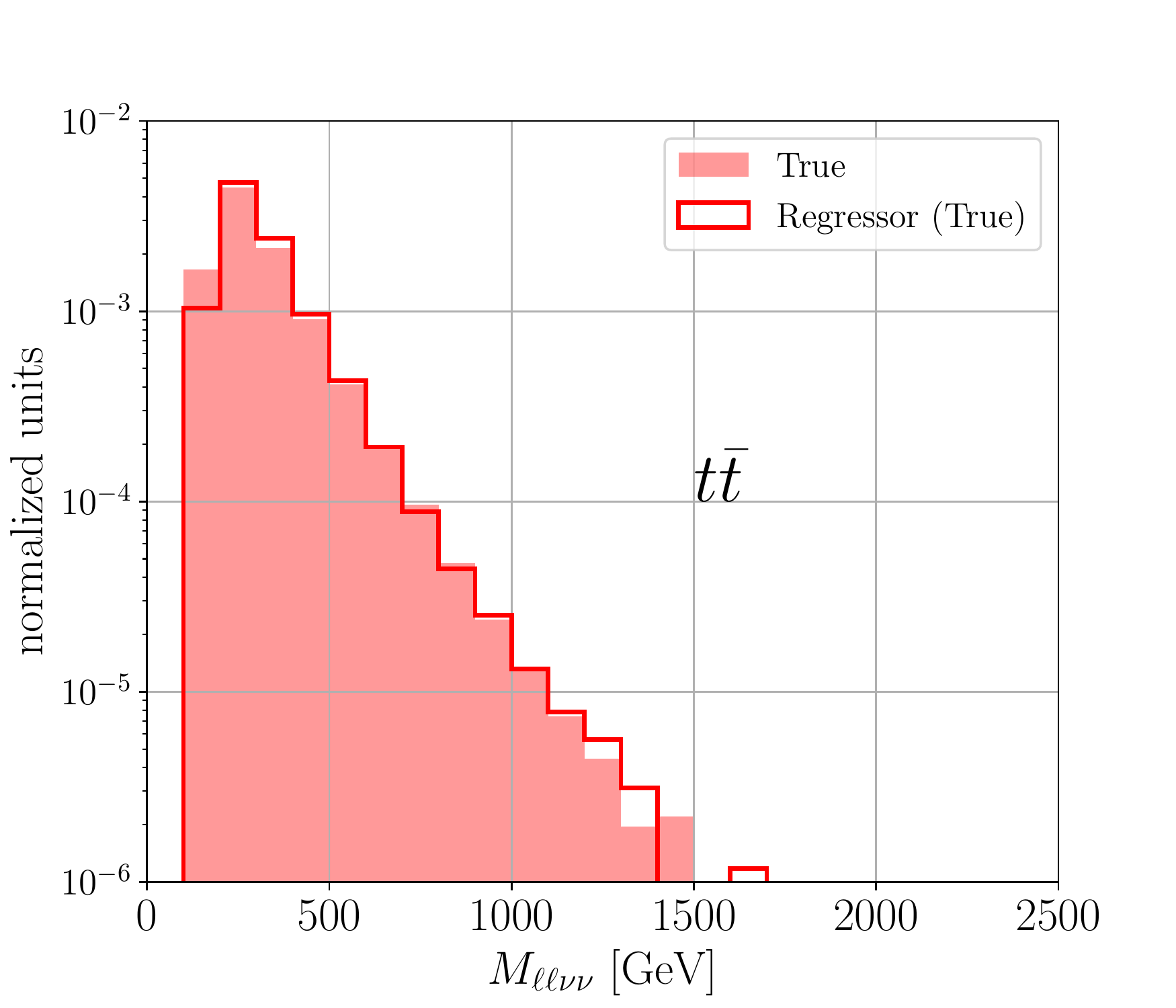}
     \caption{The true (shaded areas) and predicted (solid lines) $\mllvv$ distributions for the 2 TeV Higgs (upper left), $WW$ (upper right), $ZZ(\gamma^*)$ (lower left) and $t\bar{t}$ background (lower right). The regression is based on true samples, in this case.} 
     \label{fig:mllvv_regressor}
 \end{figure}
 %

 The number of  PCA dimensions where the original data representation is projected onto showed a more significant variation. While for the $ZZ$ background and the 1 TeV Higgs, the smaller MSE could be reached with just a one-dimensional latent space, the $WW$ background performed better in a two-dimensional PCA space, the $t\bar{t}$ and a 1.5 TeV Higgs with 3 PCA dimensions, and the 2 TeV Higgs with 6 PCA dimensions. We thus observe that as the particles get heavier, the higher should be the dimension of the PCA space. The choice of the distance metrics has no impact on the performance of the algorithms once the uniform weights performed better than the weighted option in all experiments. It means that the prediction is a simple arithmetic mean of the nearest neighbors of a given point projected on the principal component space of the events. We also tested non-linear transformations to the latent space as TSNE, but with marginal gains at the cost of much longer computation time.


 In Fig.~\eqref{fig:mllvv_regressor}, we display the true and the predicted $\mllvv$ masses for a 2 TeV Higgs boson, with $\Gamma_H/m_H=10$\%, and the $WW$, $ZZ(\gamma^*)$, and $t\bar{t}$ backgrounds. As we see, the regressors work very well for each class of events. The binning of the distributions also affects the quality of regression. We found that the mean square error between the true and predicted distribution gets larger as the bin widths get smaller as expected. It is easier to predict in which bin an event will fall when it is wide. We checked that the width of the resonances affects too little the accuracy of the regression from $\Gamma_H/m_H=1$\% up to 10\%.

\subsection{Pre-regression classification}
\label{sec:pre}

The construed $\mllvv$ regressor of a given class can predict the target distribution of events that pertain to that class exclusively. If one feeds a background regressor with a signal event, for instance, the background regressor will find the target value of the background distribution, which is closer to the signal event. In order to predict the classes' targets correctly, we need first to predict the classes as accurately as possible. We also need to know the mass of the resonance.
 
 The classification of events was performed with neural networks (NN)~\cite{Goodfellow-et-al-2016,2019} based on the same features used for regression. We took 1.5 million signals and 4 million backgrounds events to tune, train and test the algorithms. As we will discuss later, this body of data was further split to independently adjust, train, and test a second neural network; that is why we need such a large number of simulations.
 
 Beside the kinematic variables described in the previous section, we also constructed a new one that we describe now. 
 \begin{figure}[t!]
     \centering
     \includegraphics[scale=0.45]{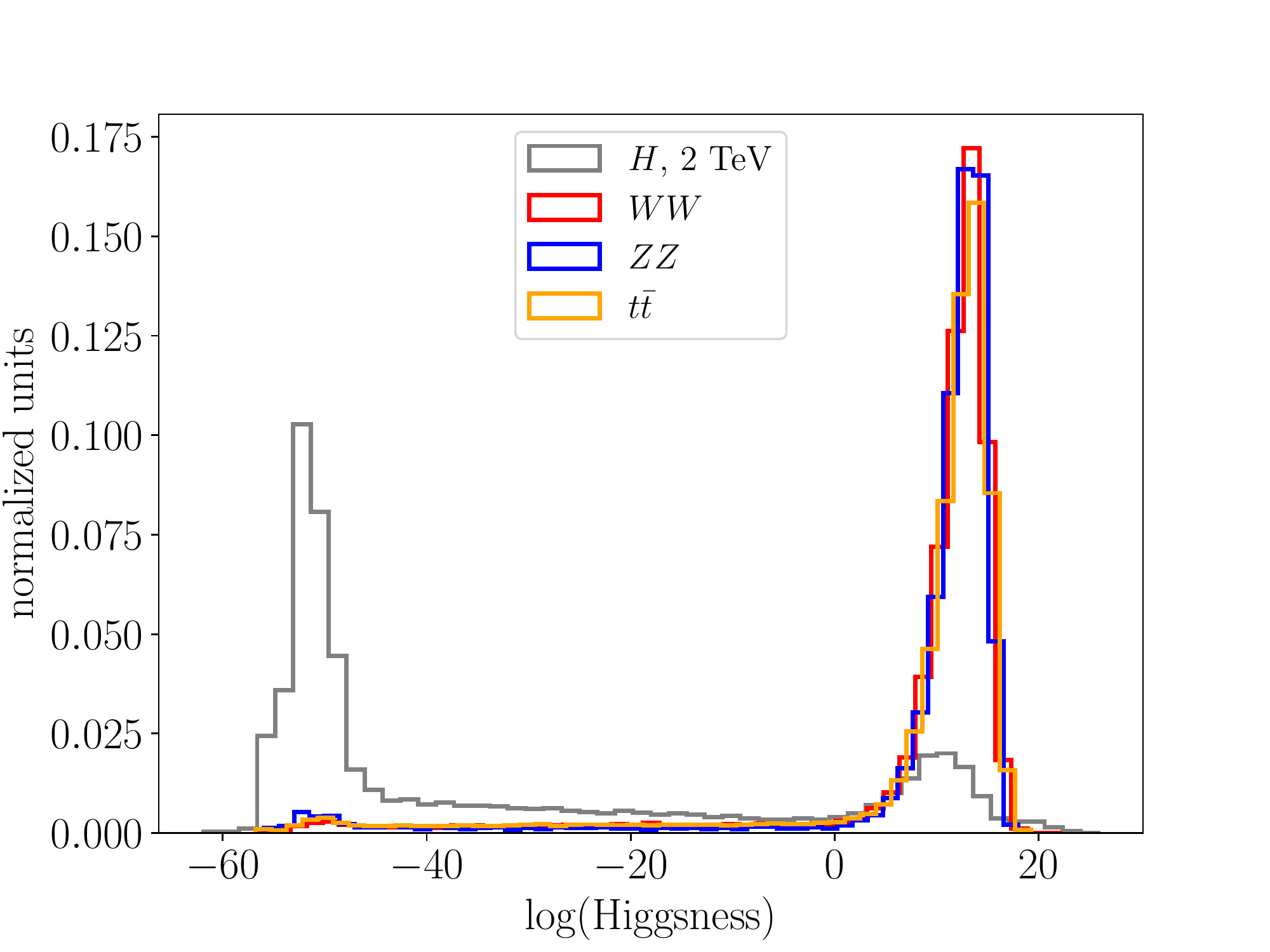}
     \caption{The logarithm of the Higgsness variable defined in Eq.~\eqref{eq:higgsness} for a 2 TeV Higgs boson and its SM backgrounds.}
     \label{fig:Higgsness}
 \end{figure}
 In Ref.~\cite{Kim:2018cxf}, a kinematic variable, called {\it Higgsness}~\footnote{Topness was another variable proposed in this reference to better tag $t\bar{t}$ events.}, is introduced to denounce the presence of a SM Higgs boson decaying to $W^\pm W^{*\mp}\to \ell^+\ell^{-\prime} + \nu_\ell\bar{\nu}_{\ell^\prime}$. The idea is to search for the neutrinos 4-momenta of an event which minimize
 \begin{equation}
     \hbox{Higgsness}\equiv \underset{p_\nu, p_{\bar{\nu}}}{\mathrm{argmin}}\left[\frac{(M_{\ell^+\ell^-\nu\bar{\nu}}^2 - m_H^2)^2}{\delta_H^4} + \min\left(\frac{(M^2_{\ell^+\nu}-m_W^2)^2}{\delta_W^4},\; \frac{(M^2_{\ell^-\bar{\nu}}-m_W^2)^2}{\delta_W^4}\right)\right],
     \label{eq:higgsness}
 \end{equation}
 where $\delta_H$ and $\delta_W$, in principle, represent experimental uncertainties, but for our purposes, they can be treated as free parameters. In fact, the value of these parameters matters for the Higgsness distributions, and we adjust them for maximum discernment among the classes. 
 
  In Fig.~\eqref{fig:Higgsness}, we show the distribution of the logarithm of Higgsness for a 2 TeV Higgs boson and the $WW$, $ZZ$ and $t\bar{t}$ backgrounds. We used a simplex algorithm from \texttt{SciPy}~\cite{2020SciPy-NMeth} to search for the minimum of the Higgsness variable. As expected, Higgsness is very small for signal events, while it is much bigger for a background event. 
 %
 \begin{table}[b!]
 \centering
 \begin{tabular}{c|c|c|c}
 \hline
    Hyperparameter/architecture  &  1 TeV & 1.5 TeV & 2 TeV\\
    \hline\hline
    L2 regularization  & $5.5\times 10^{-6}$ & $4.7\times 10^{-8}$ & $3.3\times 10^{-8}$ \\
    initial learning rate & $2\times 10^{-3}$ & $8.2\times 10^{-3}$ & $9.6\times 10^{-3}$ \\
    batch size & 160 & 160 & 128 \\
    dropout rate & 0.03 & -- & -- \\
    weight initialization & normal & uniform & uniform \\
    layer activation & $\tanh$ & ReLU & $\tanh$ \\
    numbers of layers and neurons & (64,32,16,8) & (160,80,40,20) & (128,64,32,16) \\
    total of parameters & 4188 & 20544 & 13748 \\
    \hline\hline
 \end{tabular}
\caption{Hyperparameters and architecture of the neural network classifiers to clean Higgs boson signals of 1, 1.5 and 2 TeV masses from its SM backgrounds. No dropout layers were needed in the 1.5 and 2 TeV cases.}
\label{tab:hyper}
 \end{table}

 We used \texttt{Keras}~\cite{chollet2015keras} with the \texttt{Tensorflow2.0}~\cite{tensorflow2015-whitepaper} backend to build multiclass NN classifiers. The tuning of the architecture and hyperparameters were done with \texttt{Hyperopt}~\cite{hyperopt}. An initial learning rate was adjusted following a schedule halving every ten epochs. The training was halted if no improvements on the validation loss were observed over 20 epochs or a maximum of 100 epochs was reached. The model delivering the smaller validation loss was selected during the training phase. We trained different models to identify the Higgs boson of 1, 1.5, and 2 TeV masses. The hyperparameters and the neural network architectures are shown in Table~\eqref{tab:hyper}. We split the data in proportion to 70\%, 20\%, and 10\% for training, testing, and validation of the classifiers, respectively. What we learn is that the 1 TeV Higgs bosons need a more regularized model to be discerned from backgrounds in the test samples compared to heavier masses with a stronger L2 regularization, dropout layers, and a less complex architecture. It reflects the fact that it is harder to separate lighter resonances from the SM backgrounds.
 %
 \begin{figure}[t!]
     \centering
     \includegraphics[scale=0.52]{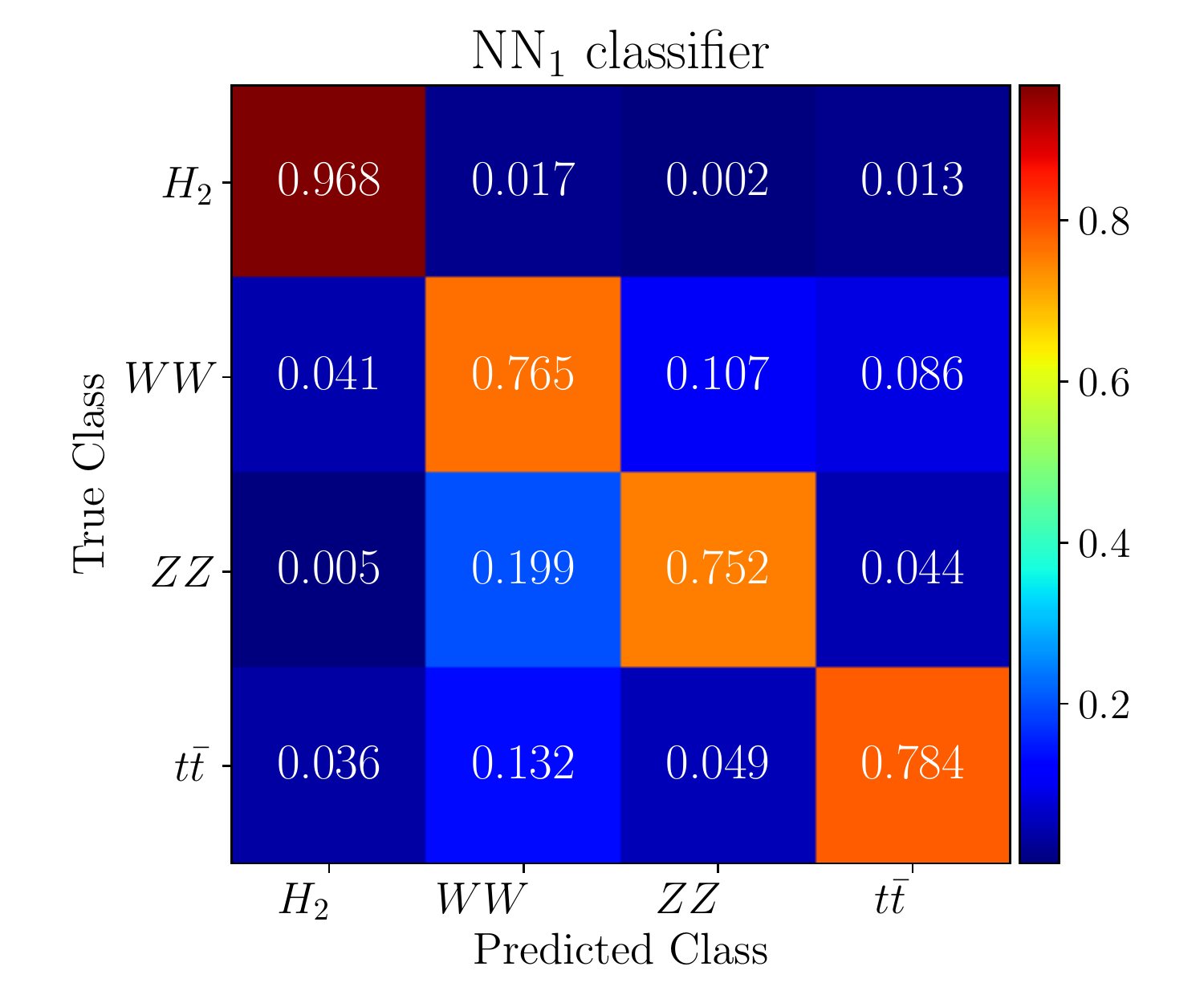}
     \includegraphics[scale=0.45]{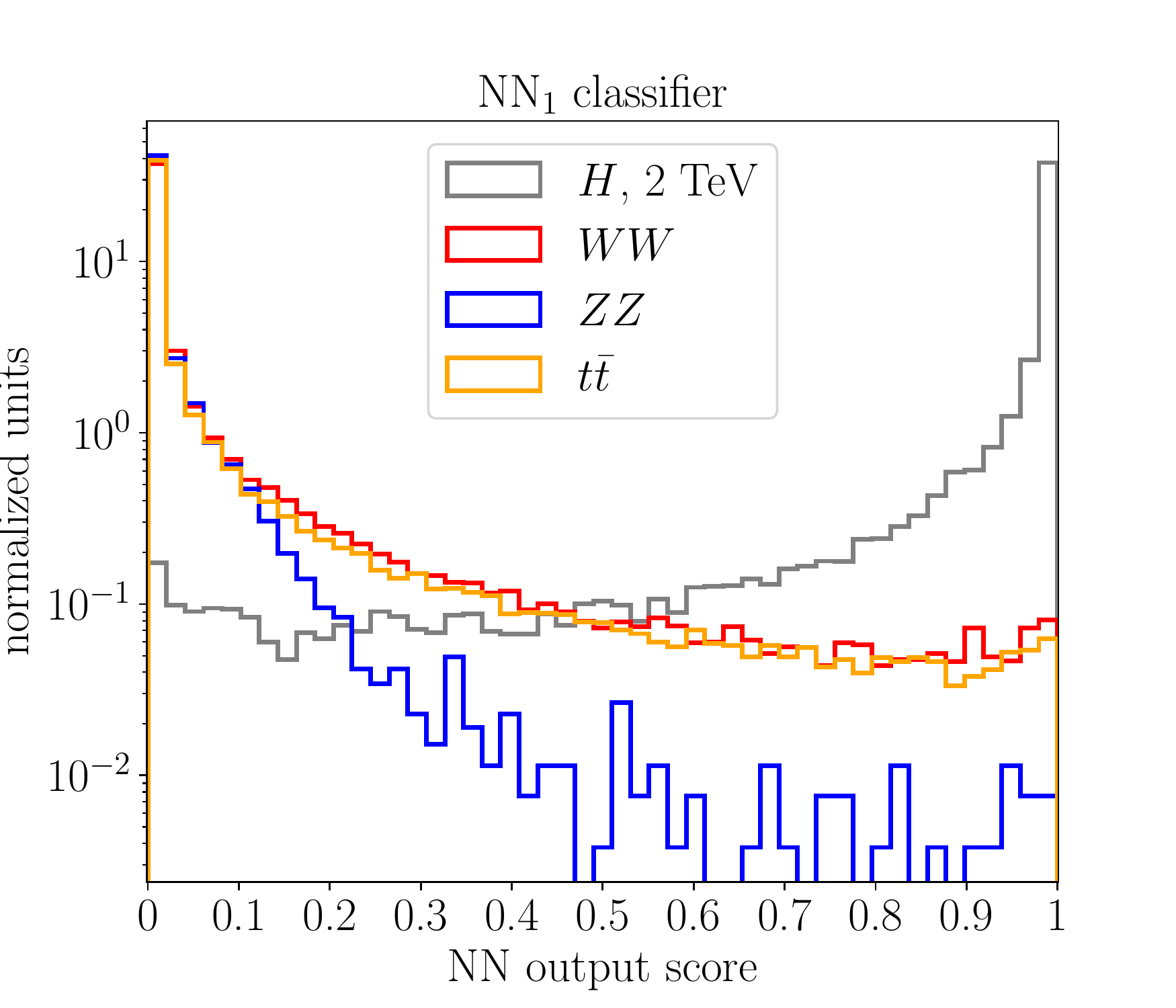}
     \caption{Confusion matrix of the classification and the output scores of the neural network before $\mllvv$ regression at the left and right panels, respectively, for a 2 TeV Higgs boson and its main SM backgrounds.}
     \label{fig:cm1}
 \end{figure}

 In Fig.~\eqref{fig:cm1}, we display the confusion matrix of the NN classifier (let's called it NN$_1$) trained to recognize the signals of a broad 2 TeV Higgs boson resonance, with $\Gamma_H=200$ GeV, against the $WW$, $ZZ$ and $t\bar{t}$ events at the left panel, and the output scores of each class at the right panel. As expected, $WW$ and $t\bar{t}\to W^+W^- +b\bar{b}$ events are more frequently mistagged by the classifier, with 13(11)\% of the $t\bar{t}$($WW$) sample tagged as a $WW$($t\bar{t}$) event. Looking at Fig.~\eqref{fig:distr}, we indeed see that $WW$ and $t\bar{t}$ events look similar once the decay of the top quark to a $W$ boson plus a $b$-jet. On the other hand, around 1/3 of all $t\bar{t}$ events have no tagged $b$-jets, the most important discriminant against $W$ pair production. This similarity is summarized in the right panel of Fig.~\eqref{fig:cm1} where we see that the scores distributions of $WW$ and $t\bar{t}$ events overlap.
 
 However, the most mistagged class is $ZZ$, where 19\% of the sample is classified as a $WW$ event. On the other hand, only 3\% of signal events are wrongly assigned to background classes. The same behaviour was observed for the other two mass values. This somewhat large misidentification of $ZZ(\gamma^*)$ events might be explained by the introduction of Higgsness as a feature of the dataset. As we see in Fig.~\eqref{fig:Higgsness}, while being very powerful to discern the signals, Higgsness is very similar for background classes. Withdrawing Higgsness from the data representation decreases the true positive rate of the 2 TeV Higgs boson from 97 to 94\%, while also decreasing the proportion of $ZZ$ events to be labeled as $WW$ events from 19 to 6\%. Apparently, singling out signal events with Higgsness make the background classes less discernible among themselves.

 
 %
  \begin{figure}[t!]
     \centering
     \includegraphics[scale=0.4]{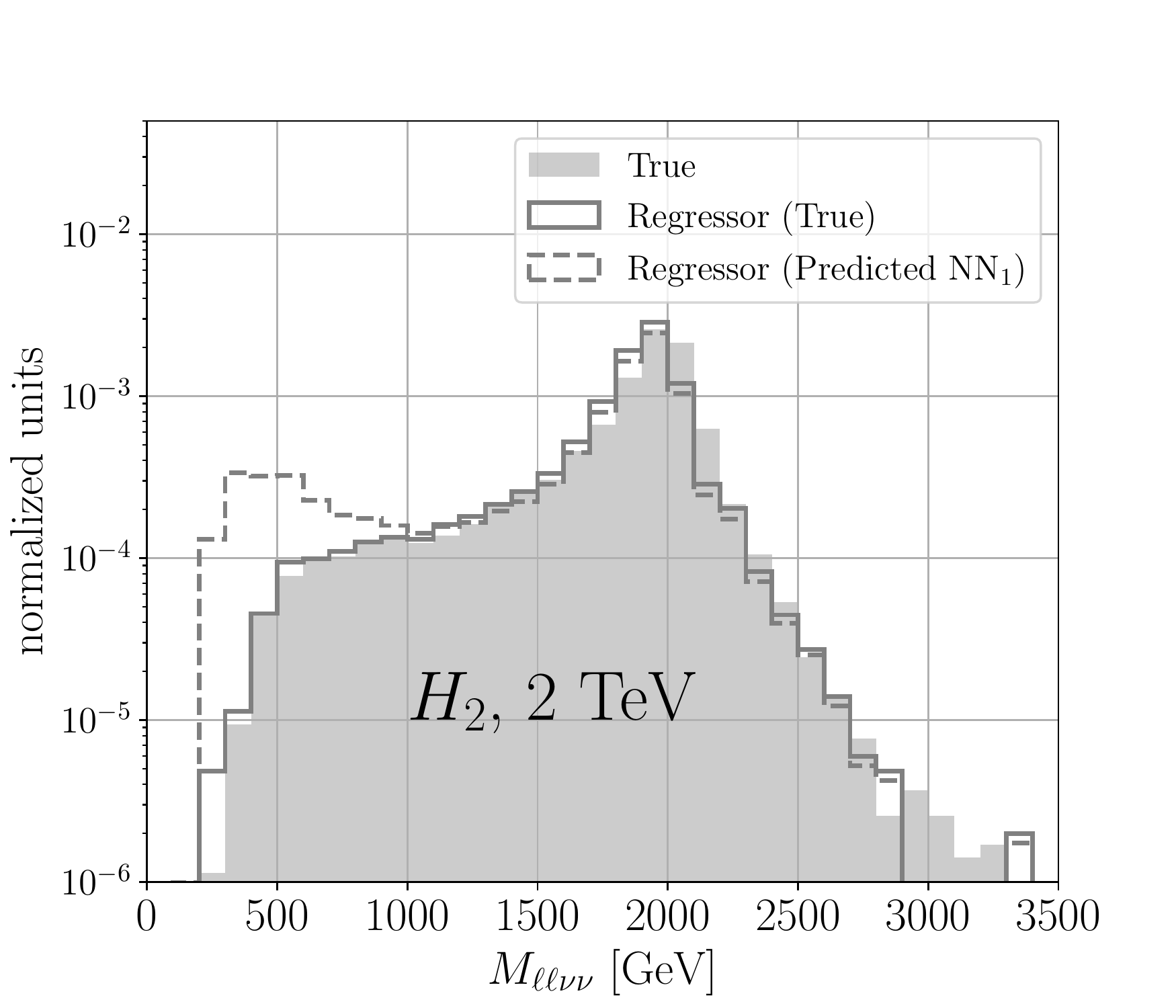}
     \includegraphics[scale=0.4]{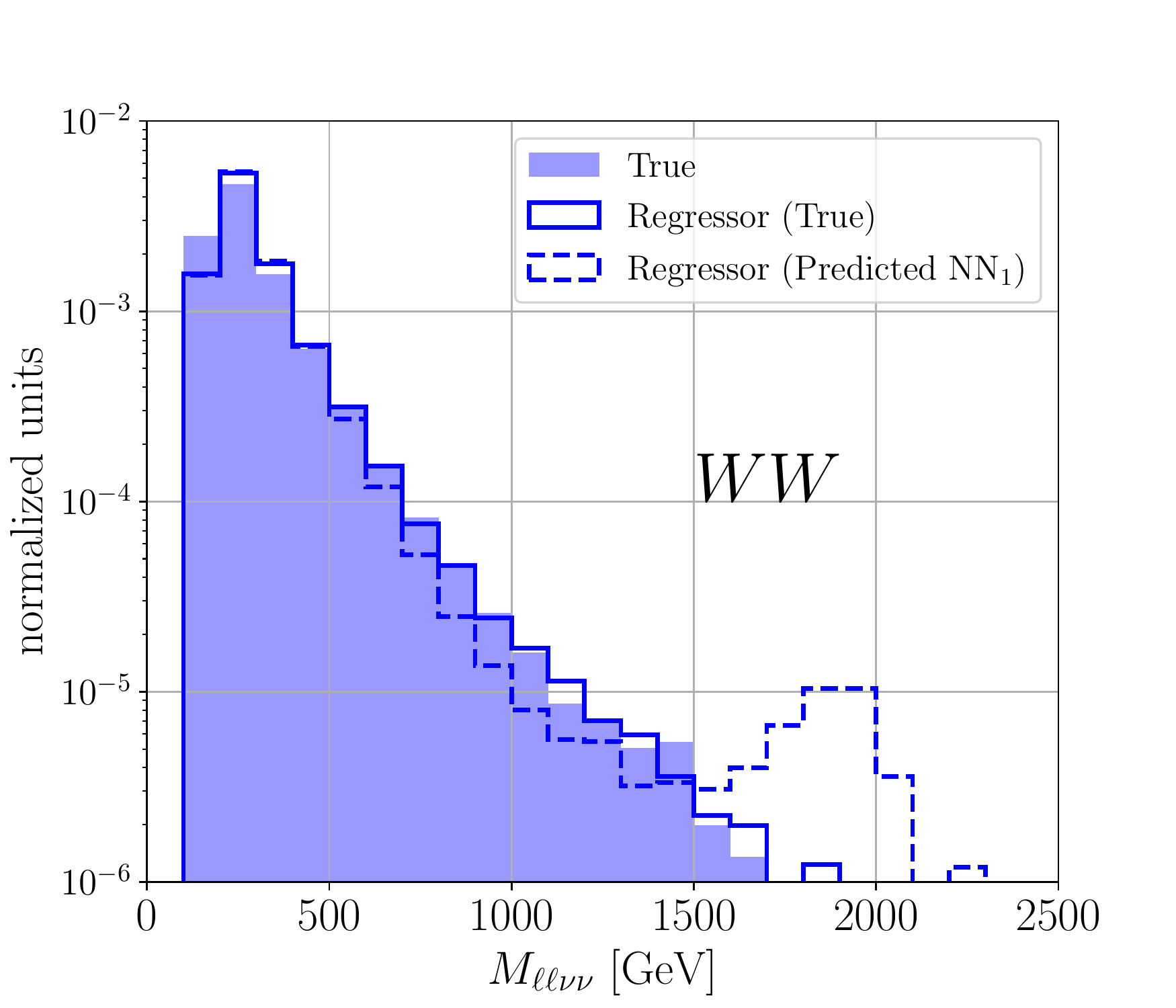}\\
     \includegraphics[scale=0.4]{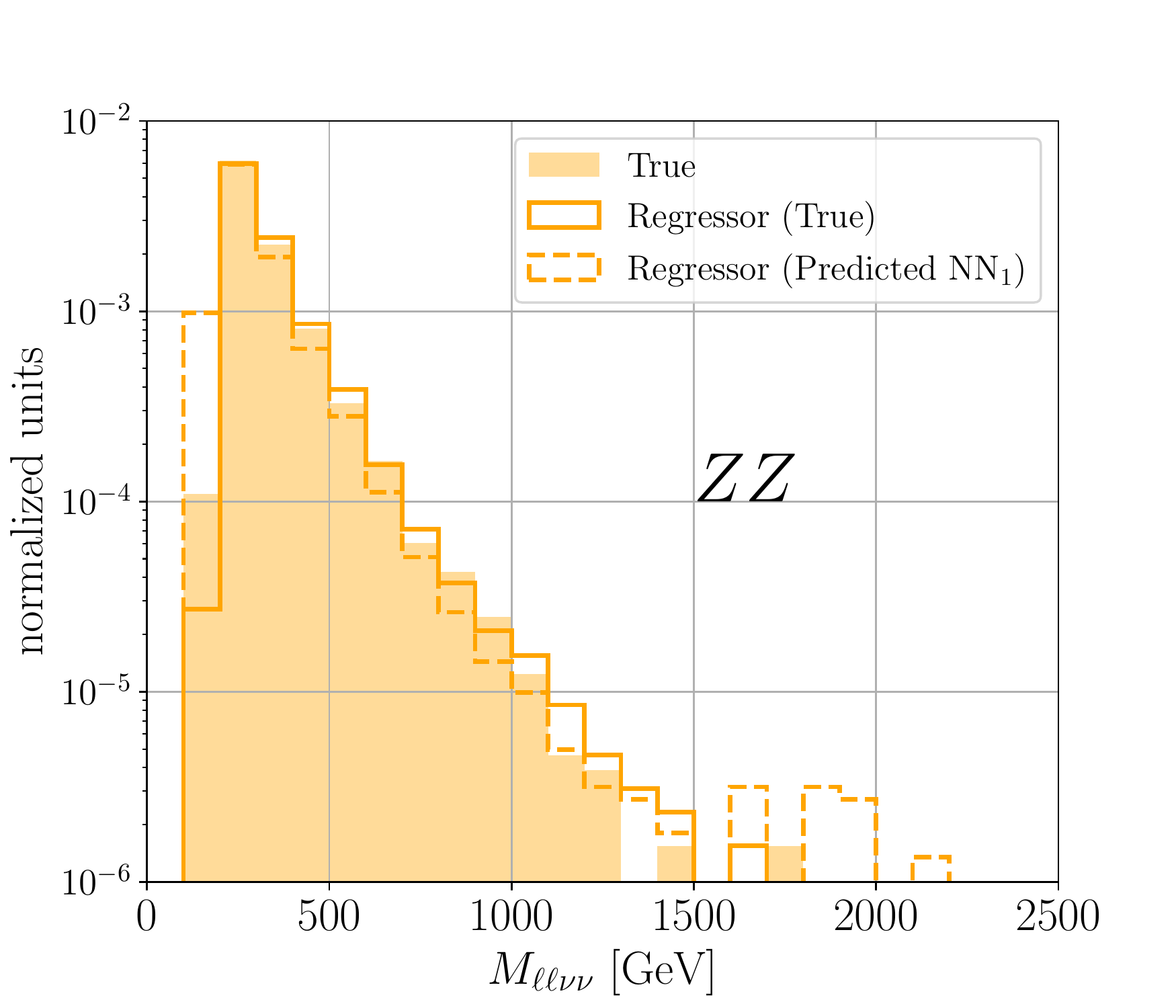}
     \includegraphics[scale=0.4]{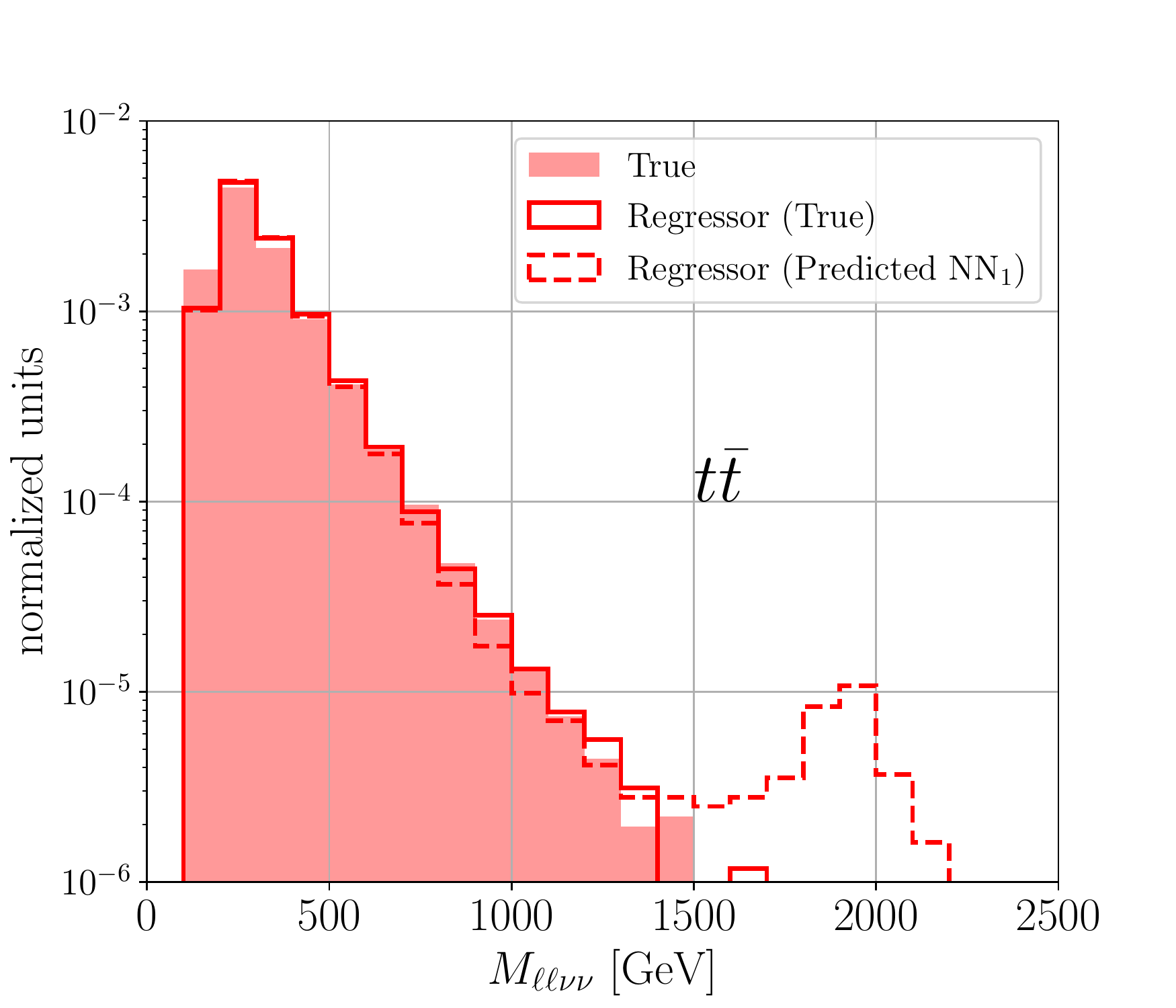}
     \caption{The true (shaded areas), regressed from true samples (solid lines), and regressed from samples identified with NN$1$ (dashed lines) $\mllvv$ distributions for the 2 TeV Higgs (upper left), $WW$ (upper right), $ZZ(\gamma^*)$ (lower left) and $t\bar{t}$ background (lower right).}
     \label{fig:regressor-1stclass}
 \end{figure}

 With the NN classifier in hand, we can reconstruct the $\mllvv$ mass of the events. We emphasize that it is necessary to know the class of the events before the regression once the target variable can only be correctly estimated when interpolated over the proper support dataset of the $k$NN algorithm. In other words, the nearest neighbors regressor does not generalize from one class to another. 
 If one presents instances never seen by the regressor, the lack of necessary correlations will result in meaningless outputs. For example, in the Higgs rest frame, the sum of the charged leptons energy and the energy of the neutrino equals the Higgs mass, $E_{\ell\ell}^*+E_{\nu\nu}^*=m_H$. In this case, the regressor can only learn the simple relation $E_{\nu\nu}^*=m_H-E_{\ell\ell}^*$ to recover the missing information from the observed one if it is trained on signal events with known $m_H$. 
 
 In Fig.~\eqref{fig:regressor-1stclass}, we show the predicted $\mllvv$ mass of the events classified by the neural network model for a 2 TeV Higgs. Again, the results for other masses and total widths are nearly the same. We note clear contamination by signal events in the tail of the distributions for $WW$ and $t\bar{t}$ events. This is expected, once 1.7 and 1.4\% of signal events are classified as $WW$ and $t\bar{t}$ events, respectively. Only 0.19\% of $H_2$ events are classified as $ZZ$ events, though and that's why we do not observe a clear peak in the tail of the $ZZ$ distribution. By its turn, 4.1 and 3.7\% of $WW$ and $t\bar{t}$ sample, respectively, is mistagged as a signal event, populating the low mass bins of the $H_2$ distribution above the true distribution. In practice, if one is interested in identifying Higgs bosons, requiring the score to be greater than 0.5 or larger is effective to mitigate the contamination of background distributions permitting a reliable estimate of backgrounds in the resonance region. However, the signal contamination is not affected much, yet, once the signal distribution contamination occurs for low mass bins, the resonance region estimate is also reliable.
 
 A way around these contaminations in order to improve the confidence in the mass estimates is presented in the next section.

\subsection{Post-regression classification and the $k$NNNN algorithm}
\label{sec:post}
\begin{figure}[t!]
     \centering
     \includegraphics[scale=0.5]{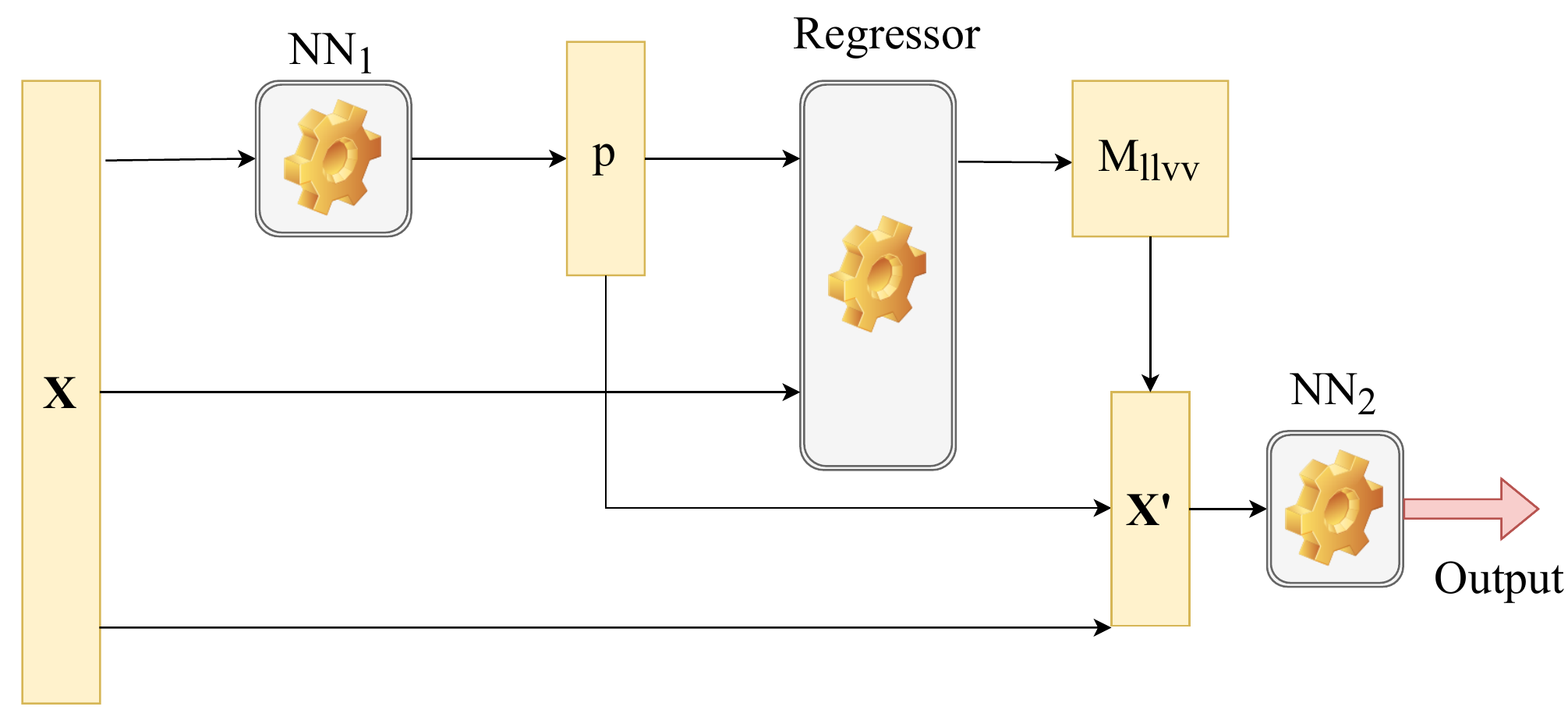}
     \caption{Flow chart of the combined classification/regression algorithm with stacking -- $k$NNNN for short. All kinematic variables described in Section~\ref{sec:reco} are passed to the Regressor except Higgsness.}
     \label{fig:flow_chart}
 \end{figure}
 %

 How can we get rid of the mistagged contamination in backgrounds and signal distributions? In Ref.~\cite{Alves:2016htj}, an ensemble of classifiers was used to boost the classification accuracy of Higgs boson events with a performance almost as good as deep neural networks~\cite{Baldi:2014kfa}. We used the same idea to boost the performance of our classifier by stacking another neural network model on the top of the first classifier described in the previous section. For a good review of ensemble methods, see~\cite{7379058}.
 
 We show a flowchart of our proposed algorithm from beginning to end in Fig.~\eqref{fig:flow_chart}. The original dataset comprising the kinematic features, $\mathbf{X}$, described in Section~\ref{sec:reco}, plus the Higgsness variable is first split into many subsets to train/validate the classifiers and the regressor. Two subsets are used to train the first classifier, depicted as NN$_1$ in Fig.~\eqref{fig:flow_chart}, and the $k$NN Regressor. In this scheme, the Regressor is fed by kinematic features, but Higgsness, and also with the output scores, $\mathbf{p}$, provided by the NN$_1$ to decide what support set should be used to calculate $\mllvv$  of a given event. After this stage, the algorithm has thus produced two important pieces of information, which are appended to $\mathbf{X}$: the scores vector, $\mathbf{p}$ and $\mllvv$, resulting in a new data representation, $\mathbf{X}^\prime$. This new representation is then used to train a second neural network, NN$_2$. Because of the combination of a $k$NN regressor with Neural Network classifiers, we call it $k$NNNN algorithm. Note that the output of NN$_2$ is the final output of the algorithm, the output of $k$NNNN itself.
 
\begin{figure}[t!]
     \centering
     \includegraphics[scale=0.52]{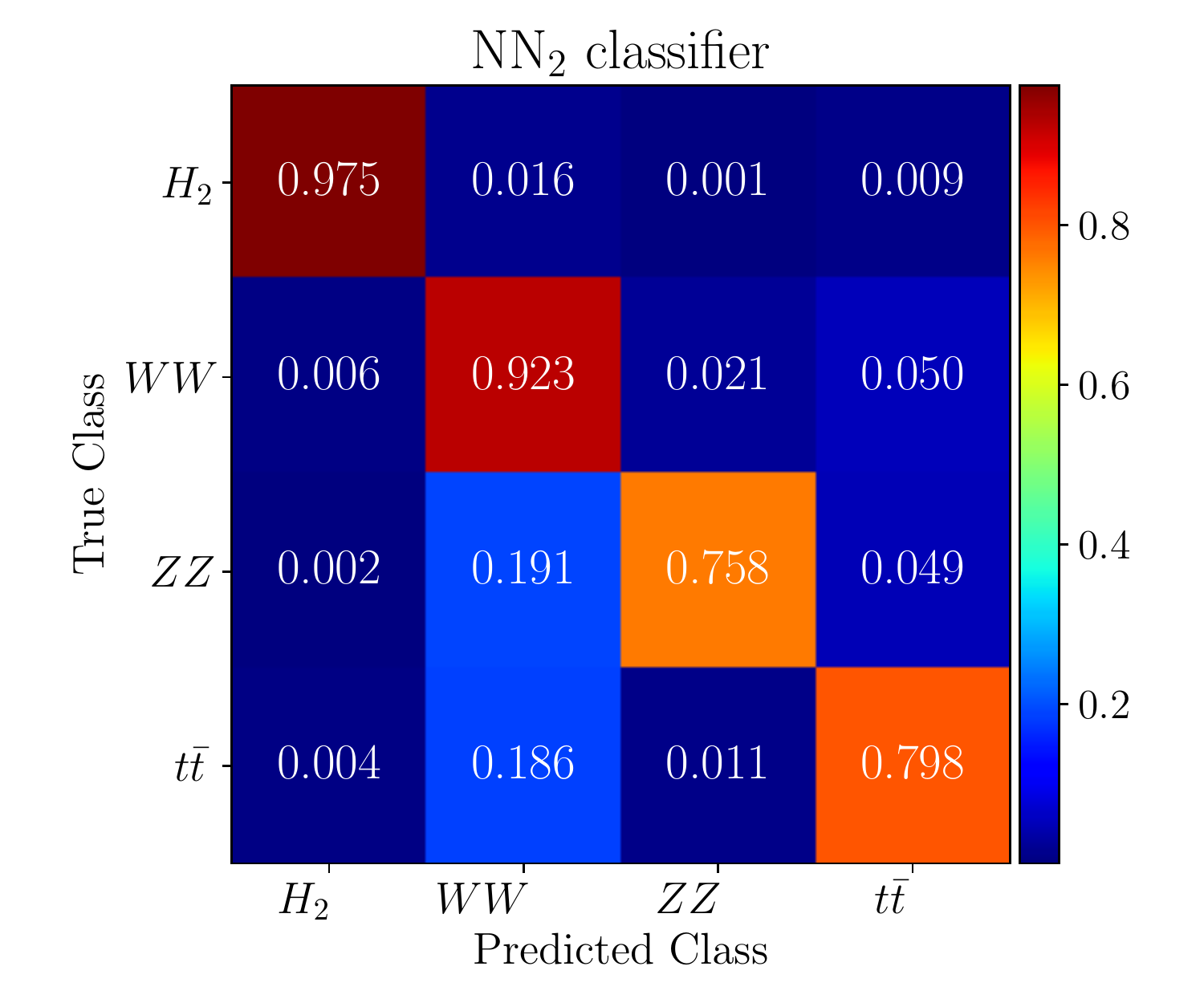}
     \includegraphics[scale=0.45]{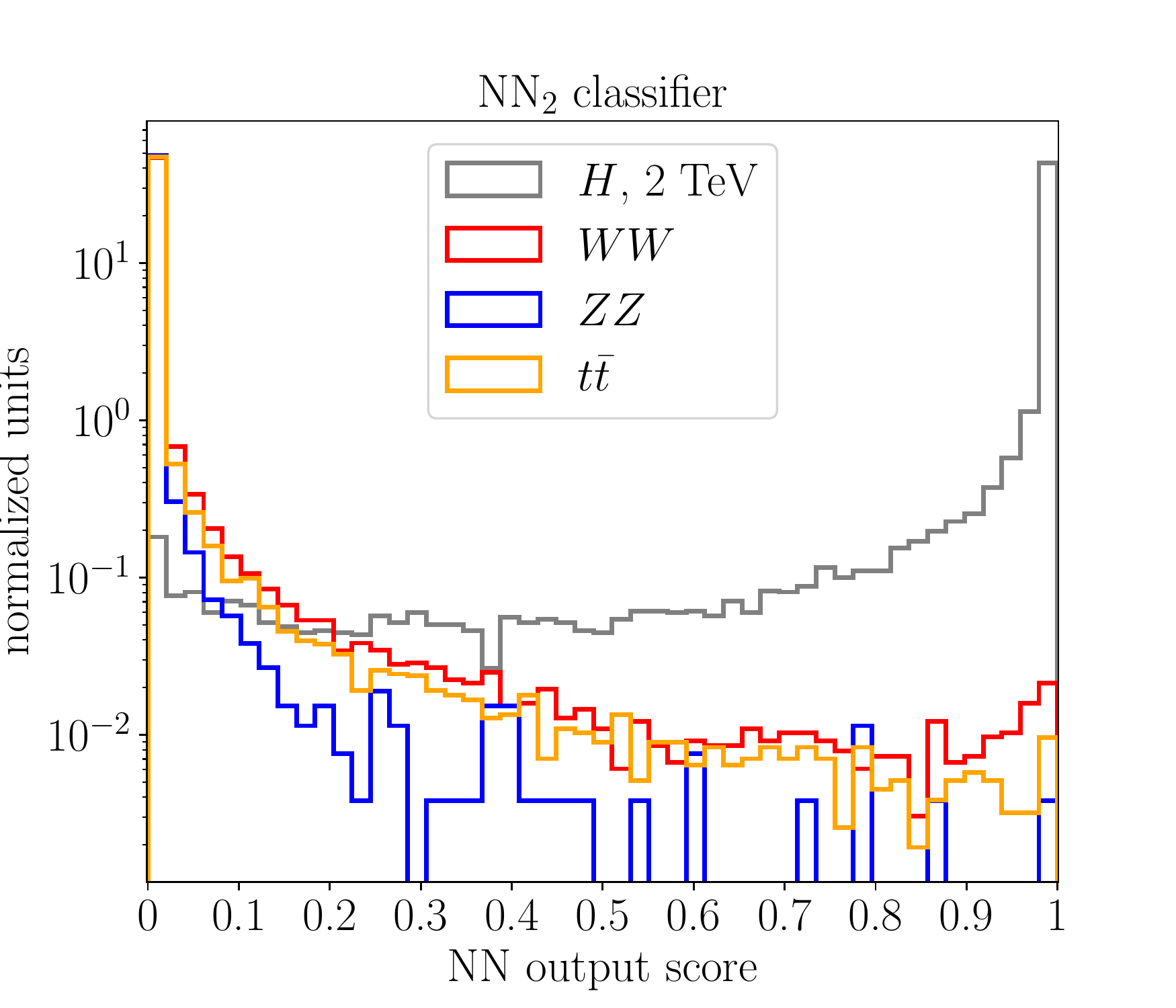}
     \caption{Confusion matrix of the classification and the output scores of the second neural network at the left and right panels, respectively, for a 2 TeV Higgs boson and its main SM backgrounds.}
     \label{fig:cm2}
 \end{figure}

 In Fig.~\eqref{fig:cm2}, we display the confusion matrix and the score outputs of the NN$_2$ classifier. The separation of the classes is improved after the second classification. To confirm that improvement, we calculate the overall accuracy and the score asymmetry defined as
 \begin{equation}
     \frac{N(\hbox{score}>0.5) - N(\hbox{score}<0.5)}{N(\hbox{score}>0.5) + N(\hbox{score}<0.5)},
     \label{eq:asym}
 \end{equation}
 where $N$ is the number of events of the class.
 
%

We also compute the positive and negative likelihood ratios, as defined in Ref.~\cite{japkowicz_shah_2011}
\begin{eqnarray}
    LR_+ &=& \frac{\hbox{Sensitivity}}{1-\hbox{Specificity}}=\frac{\hbox{Sensitivity}}{\hbox{False Positive Rate}}, \\
    LR_- &=& \frac{1-\hbox{Sensitivity}}{\hbox{Specificity}}=\frac{\hbox{False Negative Rate}}{\hbox{Specificity}}.
\end{eqnarray}

These two metrics are aimed to measure how effective a classifier is in predicting the classes in a binary problem. Sensitivity, the ratio between the number of events correctly classified as positives and the total number of events classified as positives, measures how good the classifier is in identifying the positive class, our $H_2$ events. Specificity, by its turn, is the ratio between the number of events correctly classified as negatives and the total number of events classified as negatives, our backgrounds. In order to apply these metrics, we gather all background events into a single negative class. Analogously to sensitivity, specificity measures how competent the classifier is in correctly identifying negative instances. 
 
 For the signals, $LR_+$ summarizes how many times more likely signals are correctly predicted to be signals than backgrounds are wrongly predicted to be a signal. On the other hand, $LR_-$ summarizes how many times less likely signals are wrongly predicted to be backgrounds than backgrounds events are correctly predicted to be a background. A better classifier must therefore maximize $LR_+$ and minimize $LR_-$. In the comparison of two classifiers, let's say, NN$_1$ and NN$_2$, if $LR_+(\hbox{NN}_2)>LR_+(\hbox{NN}_1)$ and $LR_-(\hbox{NN}_2)<LR_-(\hbox{NN}_1)$, then NN$_2$ is better than NN$_1$ in the confirmation of both positives  and  negatives. When the inequality of the first condition still holds but the second flips, then NN$_2$ is better than NN$_1$ in the confirmation of positive class but worse for the negative class. At the same time, if the inequality of the first condition flips but the second still holds, then NN$_2$ is worse than NN$_1$ in the confirmation of positive class but better for the negative class.
\begin{table}[b!]
  \begin{tabular}{c|c|c||c|c||c|c}
    \hline\hline
    \multirow{2}{*}{Metric} &
      \multicolumn{2}{c}{1 TeV} &
      \multicolumn{2}{c}{1.5 TeV} &
      \multicolumn{2}{c}{2 TeV} \\
    & NN$_1$ & NN$_2$ & NN$_1$ & NN$_2$ & NN$_1$ & NN$_2$ \\
    \hline\hline
    accuracy & 81.1\% & 87\% & 81.3\% & 87.6\% & 80.6\% & 87.5\% \\
    \hline
    asymmetry, $H_2$ & 0.821 & 0.933 & 0.887 & 0.944 &  0.917 & 0.941\\
    asymmetry, $WW$ & -0.903 & -0.978 & -0.933 & -0.988 & -0.939 & -0.991 \\
    asymmetry, $ZZ$ & -0.992 & -0.996 & -0.994 & -0.997 & -0.994 & -0.999 \\
    asymmetry, $t\bar{t}$ & -0.916 & -0.984 & -0.934 & -0.992 & -0.947 & -0.994 \\
    \hline
    $LR_+$ & 28.225 & 74.371 & 45.299 & 86.010 & 61.330 & 82.360 \\
    $LR_-$ & 0.603 & 0.581 & 0.589 & 0.578 & 0.582 & 0.578 \\
    \hline\hline
  \end{tabular}
  \caption{Comparison of metrics performance of models trained (NN$_1$ and NN$_2$) to identify Higgs boson of all the three masses considering in this work.}
  \label{tab:metricas}
\end{table}
 In Table~\ref{tab:metricas}, we display the accuracy, the asymmetry, and the positive and negative likelihood ratios just described for all the three Higgs boson masses investigated in our work. All metrics indicate an overall improvement of NN$_2$ over NN$_1$, but the gain in performance is more pronounced in the 1 TeV case. Lighter masses present attributes less discernible than the backgrounds, so profit more from an ensemble of classifiers that use more distinctive features like the classification scores and the $\mllvv$ mass.
 
\begin{figure}[t!]
     \centering
     \includegraphics[scale=0.35]{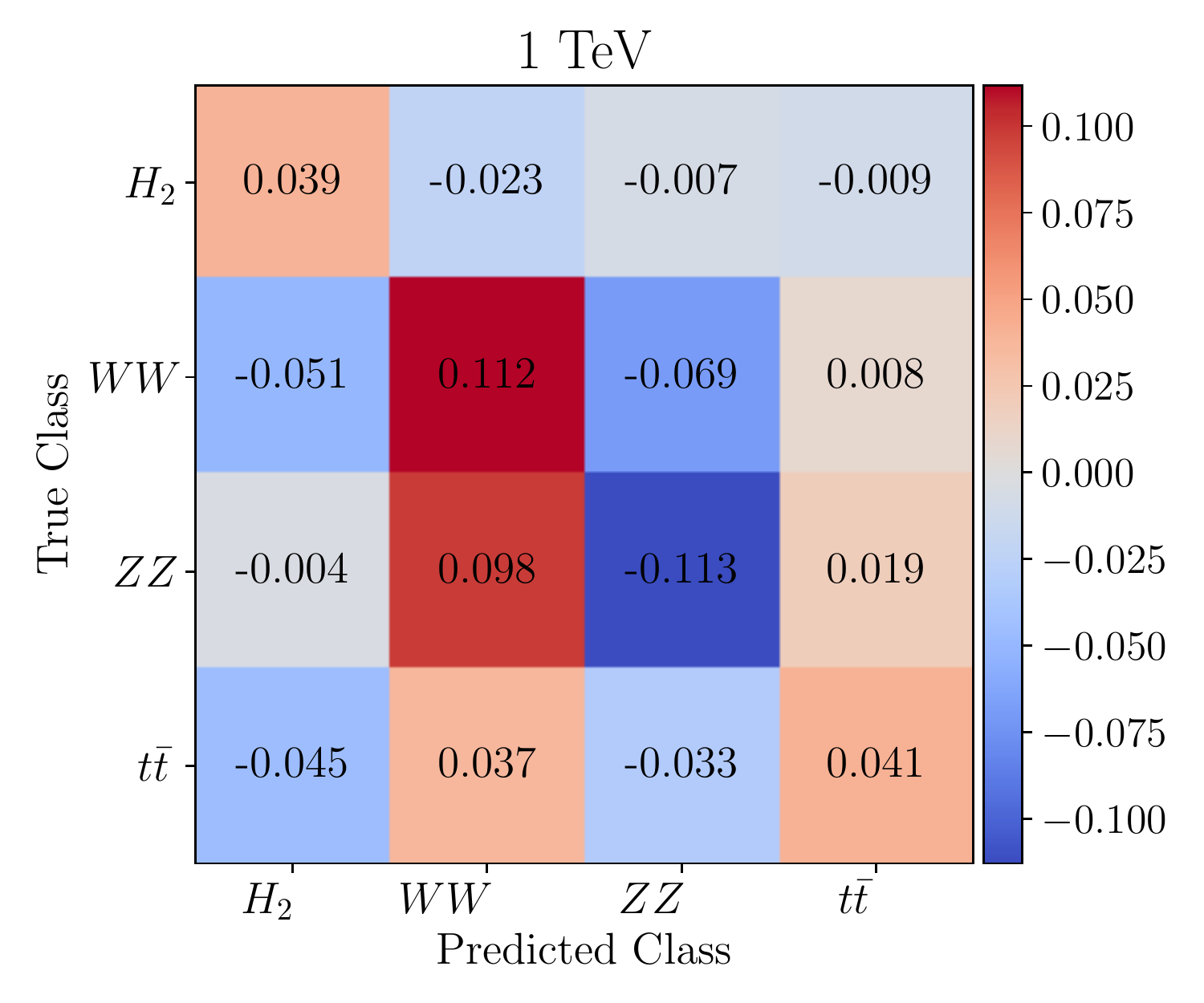}
     \includegraphics[scale=0.35]{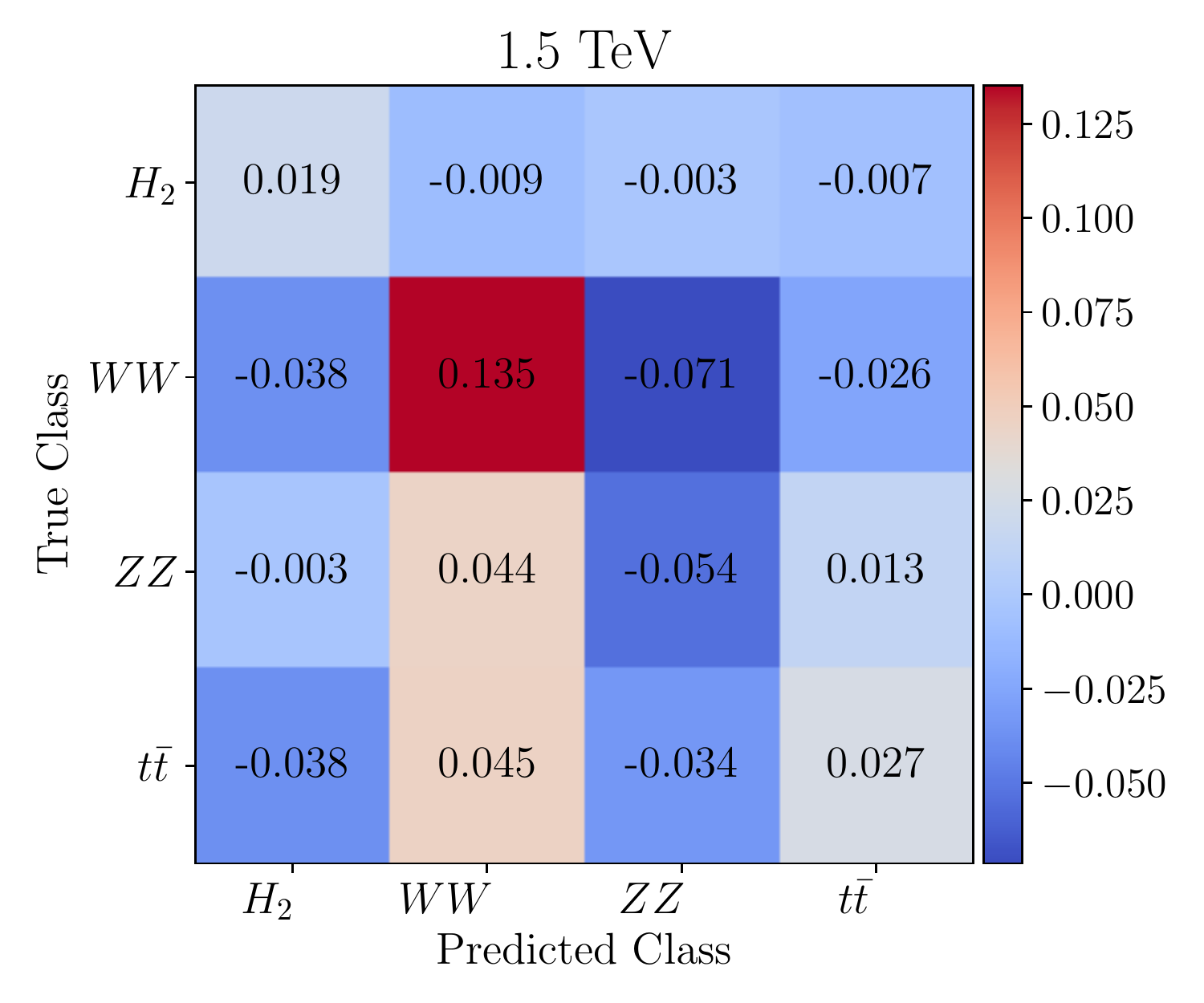}
     \includegraphics[scale=0.35]{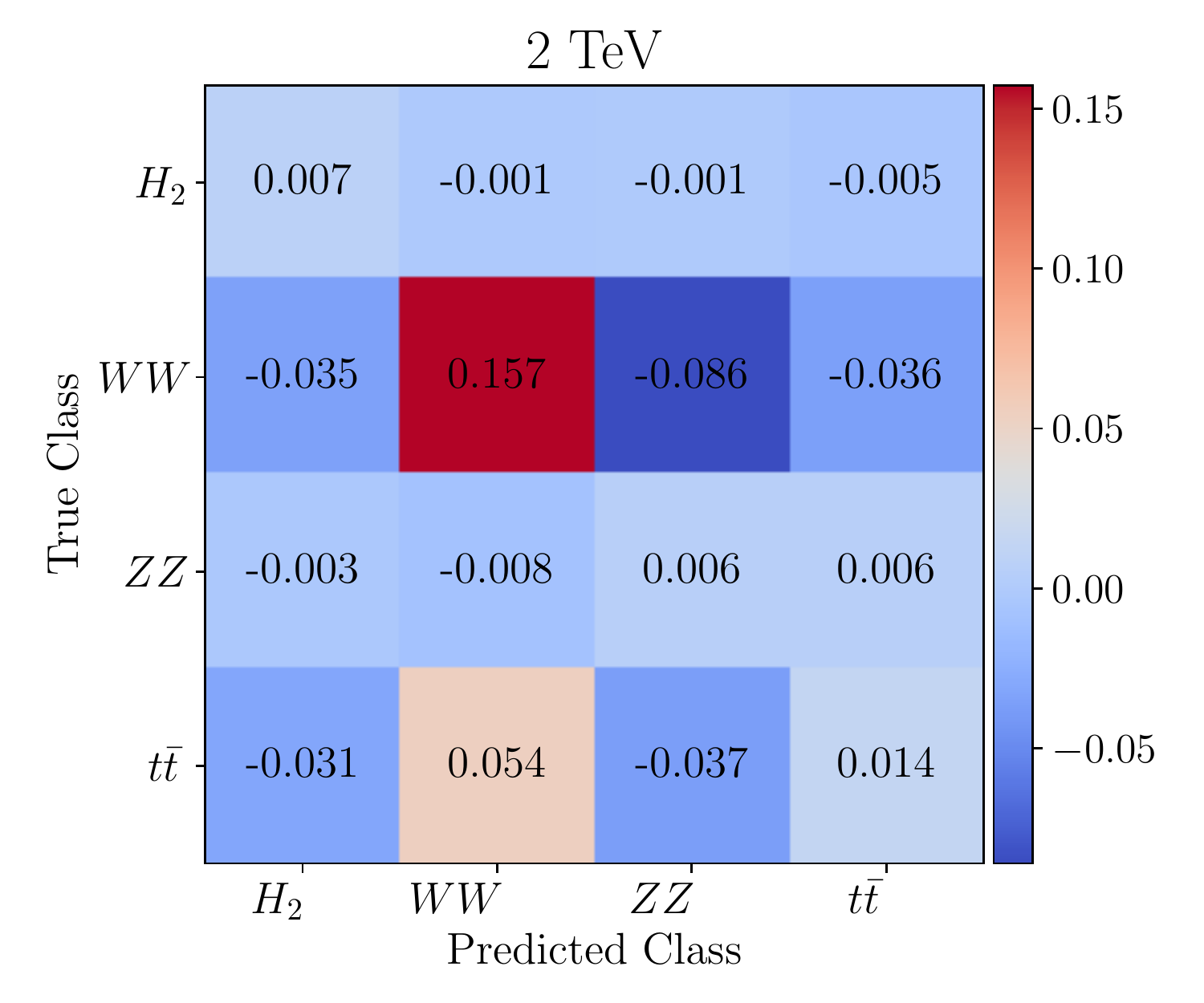}
     \caption{Confusion matrices differences (NN$_2$ $-$ NN$_1$) for models trained to separate Higgs bosons of mass 1 TeV (left), 1.5 TeV (center), and 2 TeV(right).}
     \label{fig:cm_diff}
 \end{figure}

 The improvement is more significant for the signals and the $WW$ background compared to $ZZ(\gamma^*)$ and $t\bar{t}$ events. This can be further confirmed by looking at the Fig.~\eqref{fig:cm_diff}, the difference between the confusion matrices of the NN$_1$ and NN$_2$ classifiers. First of all, we want the diagonal of Fig.~\eqref{fig:cm_diff} to be all positive, which means that NN$_2$ increases the true positive rate compared to NN$_1$. At the same time, negative non-diagonal entries mean less misclassification among classes. Overall, taking into account the results for the three Higgs masses, we see a clear improvement of NN$_2$ compared to NN$_1$. Except for the $ZZ(\gamma^*)$ class in the 1 and 1.5 TeV cases, all diagonal entries are positive, with a major improvement of $WW$ classification. Moreover, the 1 TeV signal class benefits more from NN$_2$ than the heavier masses. This is a good feature of $k$NNNN; it helps in the more difficult cases for the signals. Concerning the non-diagonal entries, we observe a clear trend -- the $ZZ(\gamma^*)$ class is more accurately identified by models whose task is to separate heavier Higgs signals. In contrast, the other classes are less confused among themselves by NN$_2$. On the other hand, the more accurate $WW$ and $ZZ(\gamma^*)$ classification comes at the cost of a slight increase in mistagging of $t\bar{t}$ events as $WW$.
 %

\begin{figure}[t!]
     \centering
     \includegraphics[scale=0.4]{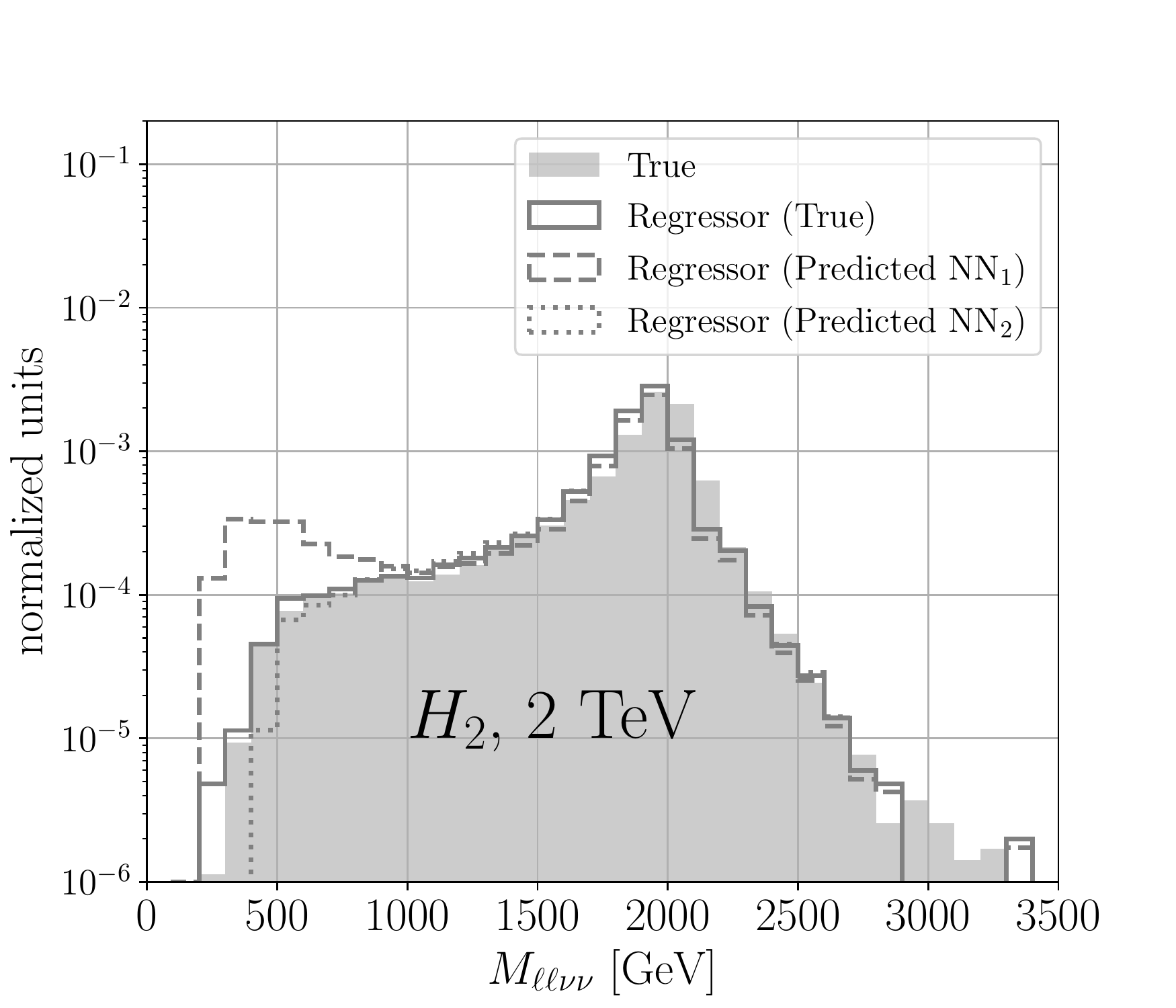}
     \includegraphics[scale=0.4]{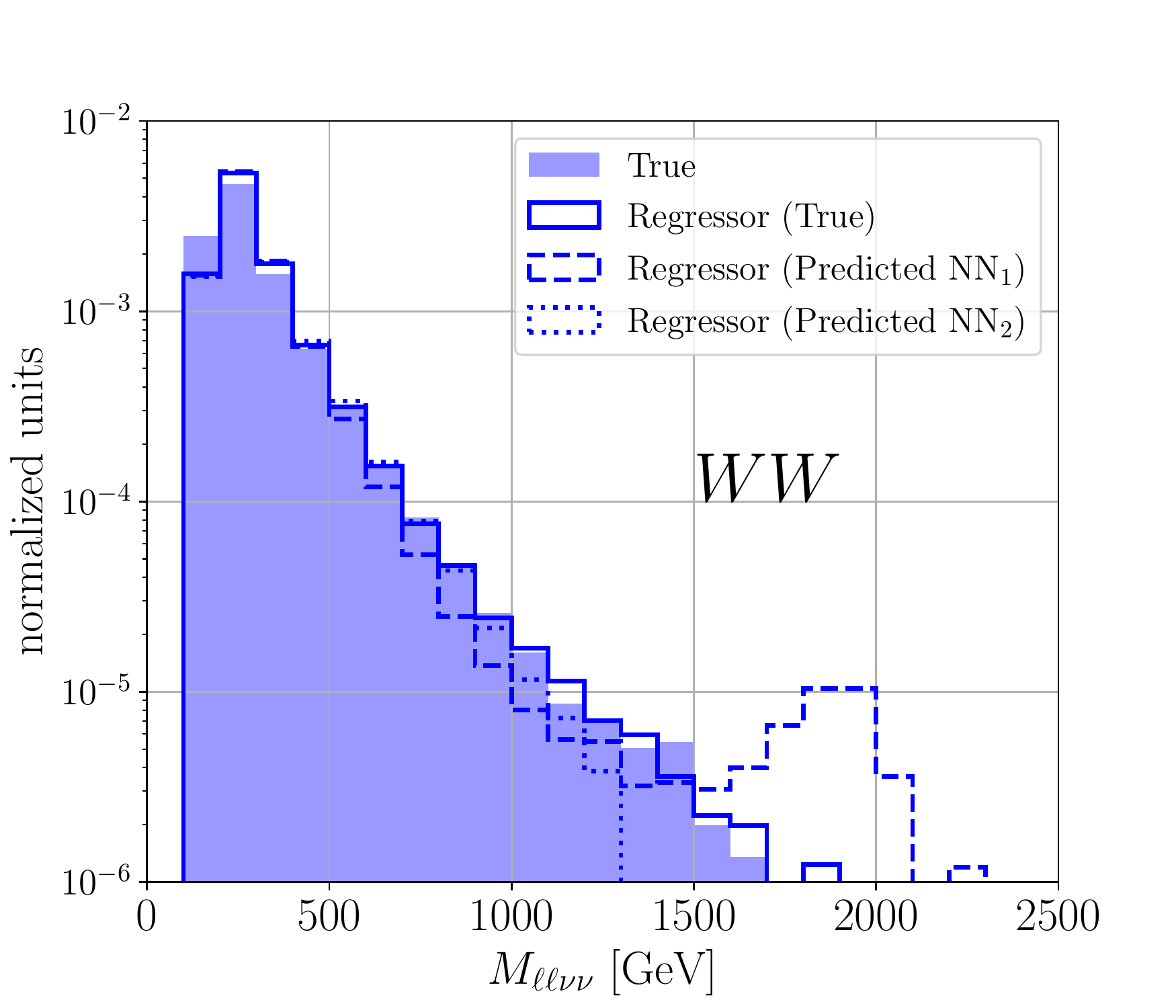}\\
     \includegraphics[scale=0.4]{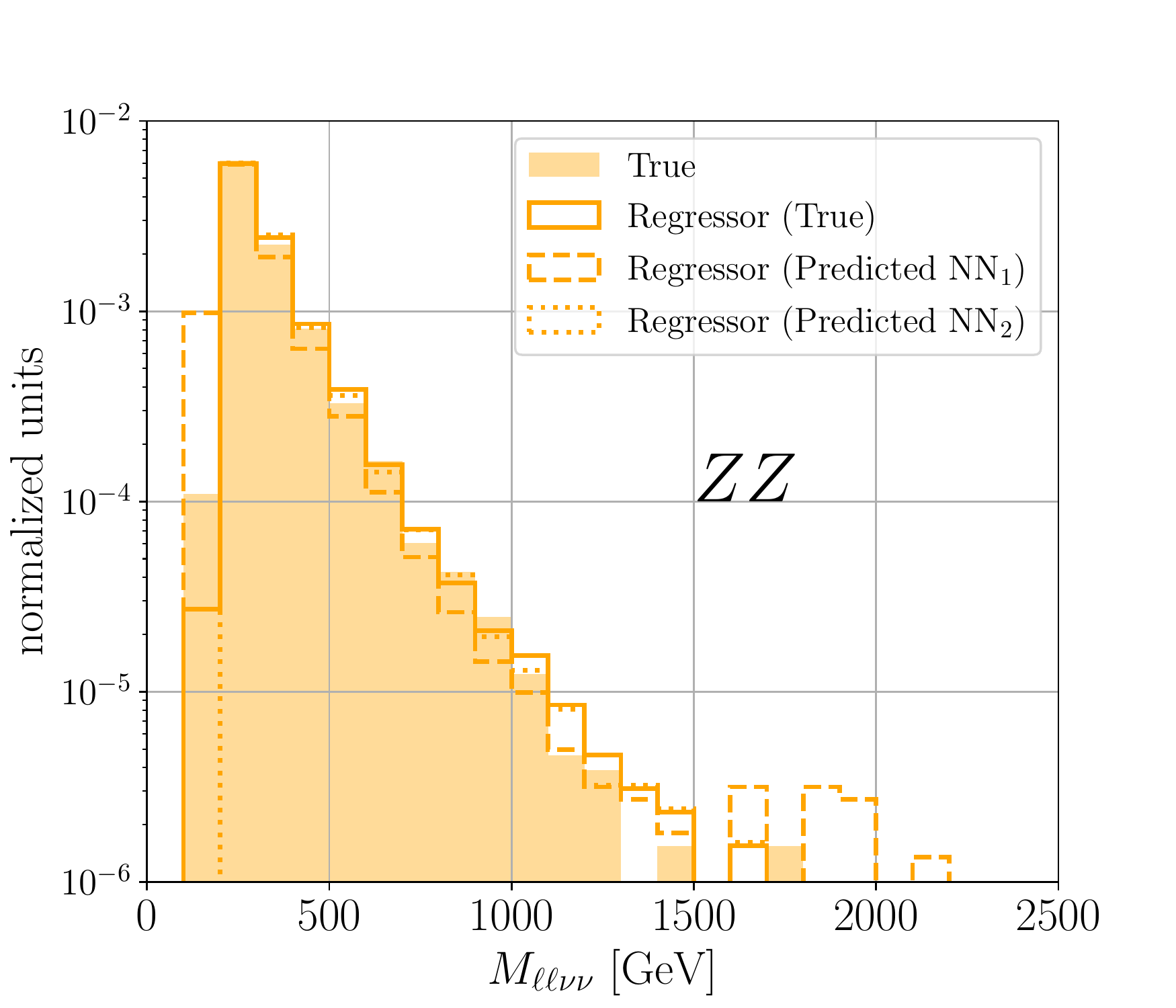}
     \includegraphics[scale=0.4]{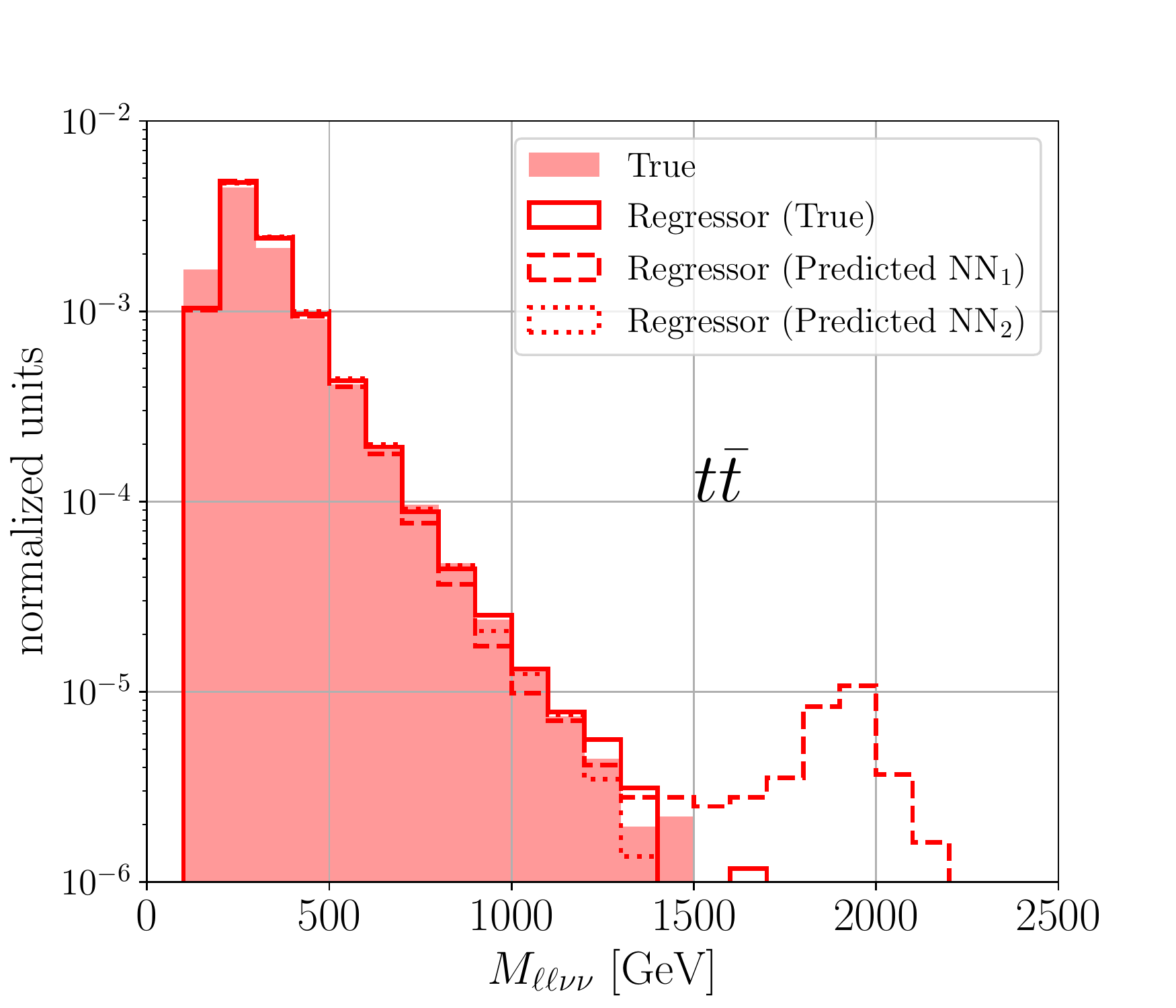}
     \caption{The true (shaded areas), regressed from true samples (solid lines), regressed from samples identified with NN$1$[NN$_2$] (dashed lines)[dotted lines] $\mllvv$ distributions for the 2 TeV Higgs (upper left), $WW$ (upper right), $ZZ(\gamma^*)$ (lower left) and $t\bar{t}$ background (lower right).}
     \label{fig:regressor-2ndclass}
 \end{figure}

 After the second classification, using the class scores of NN$_1$ and the predicted $\mllvv$ mass of the Regressor, the second neural network NN$_2$ now provides more accurate predictions to inform the Regressor which support set to use for the regression task. As an outcome, the contaminations from other classes get reduced, and the prediction of $\mllvv$ improves. We show the predicted $\ell\ell\nu\nu$ invariant mass after the second classification in Fig.~\eqref{fig:regressor-2ndclass}.

\section{Improvement of the Signal Significance}
\label{sec:results}

 Now that we have established a working algorithm to predict the $\mllvv$ mass, we want to investigate whether it is helpful to boost the statistical signal significance when employing a machine learning classifier. The signal significance is computed according to
\begin{equation}
    N_\sigma = \frac{\epsilon_{cut}^{(S)}\times N_S}{\sqrt{\sum_{i} \epsilon_{cut}^{(i)}\times N_{B_i}+(\varepsilon_B\times\sum_{i} \epsilon_{cut}^{(i)}\times N_{B_i})^2}},
    \label{eq:sigmas}
\end{equation}
where $N_S$ and $N_{B_i},i=WW,ZZ,t\bar{t}$ denote the number of signal and backgrounds events, respectively;  $\epsilon_{cut}^{(S)}$ and $\epsilon_{cut}^{(i)},i=WW,ZZ,t\bar{t}$ denote the signal and backgrounds cut efficiencies (both on kinematic variables and score outputs), respectively; finally, $\varepsilon_B$ represents a systematic uncertainty in the backgrounds rates assuming, for simplicity, a common uncertainty for all background sources.

The production cross sections of $WW$, $ZZ(\gamma^*)$, and $t\bar{t}$, at leading order are given by 102.8, 14.15 and 674.1 picobarns, respectively. The branching ratios for $W\to \ell\nu$, $Z\to\ell^+\ell^-$, $Z\to \nu\bar{\nu}$, and  $t\to bW^-$ are taken to be 10.68\%, 3.37\%, 20\%, and 100\%, respectively. Assuming the basic cuts of Eq.~\eqref{eq:cuts}, and an integrated luminosity of 500 fb$^{-1}$, we estimate $2.35\times 10^6$, $1.91\times 10^5$ and $1.53\times 10^7$ events, amounting to around $1.8\times 10^7$ background events at the 13 TeV LHC. Including NLO QCD corrections, these numbers should increase by a few tens of percent. We fix the number of signals events at 1000 for all masses for illustration purposes. The actual signal production cross section depends on the specific model of new Higgs bosons. 

As discussed previously, we are interested in showing the boost in the signal significance that our proposed algorithm is expected to produce by including the predicted $\mllvv$ mass in the data representation. Moreover, we also wish to check if the predicted masses cause an underestimation or overestimation of the statistical significance compared to what we could get if we knew the true $\mllvv$ distribution.
\begin{figure}[t!]
     \centering
     \includegraphics[scale=0.35]{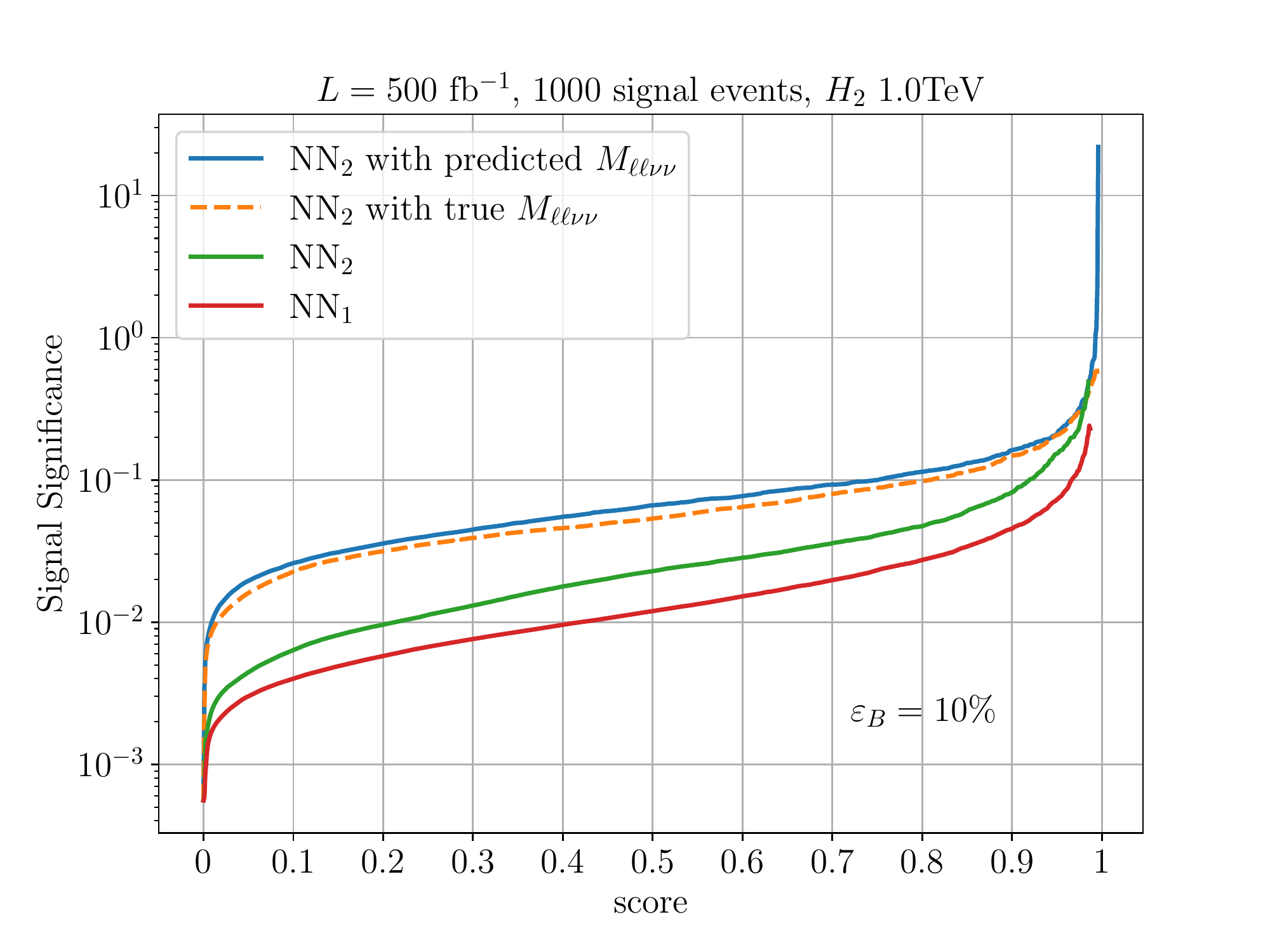}
     \includegraphics[scale=0.35]{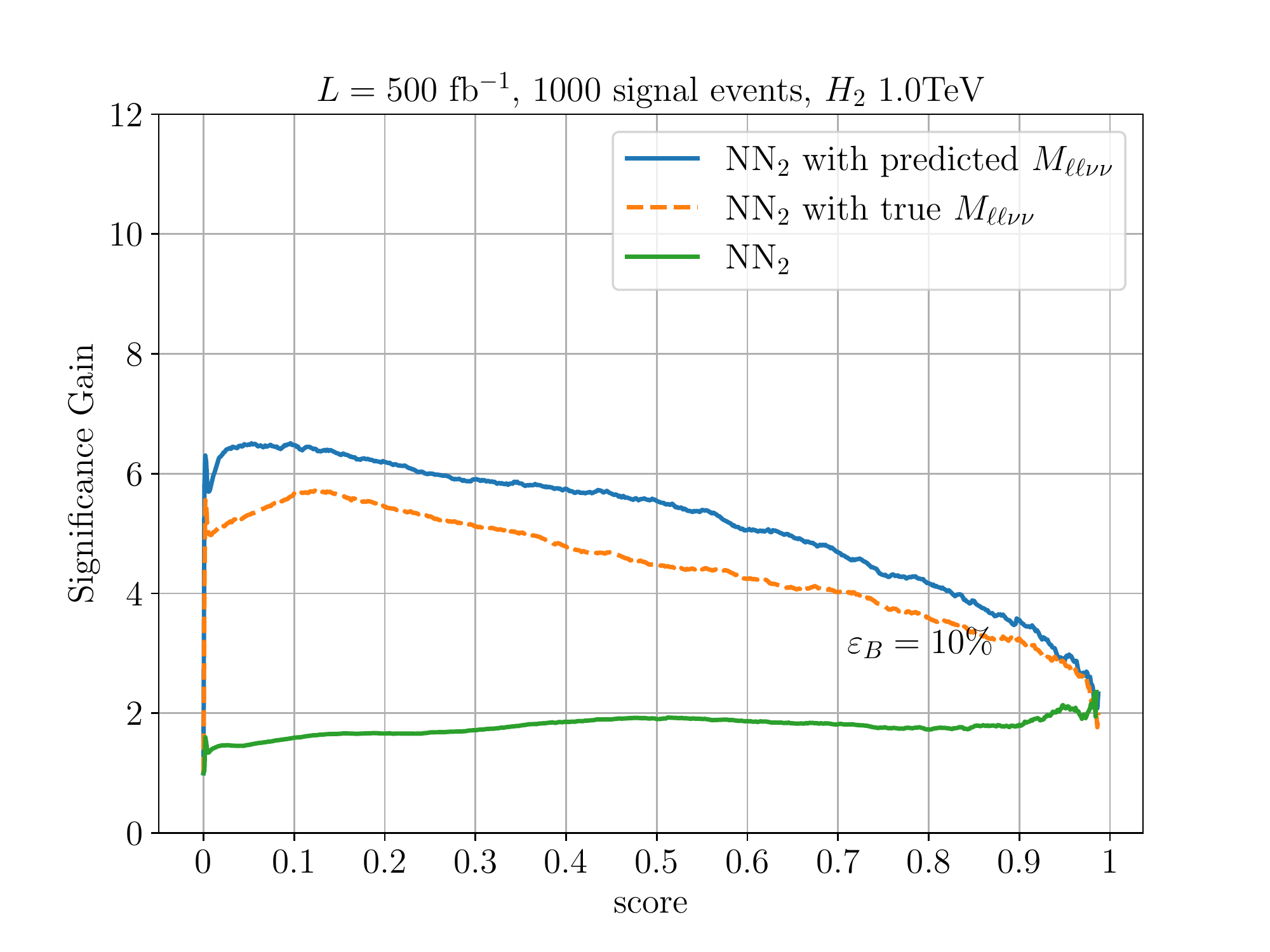}\\
     \includegraphics[scale=0.35]{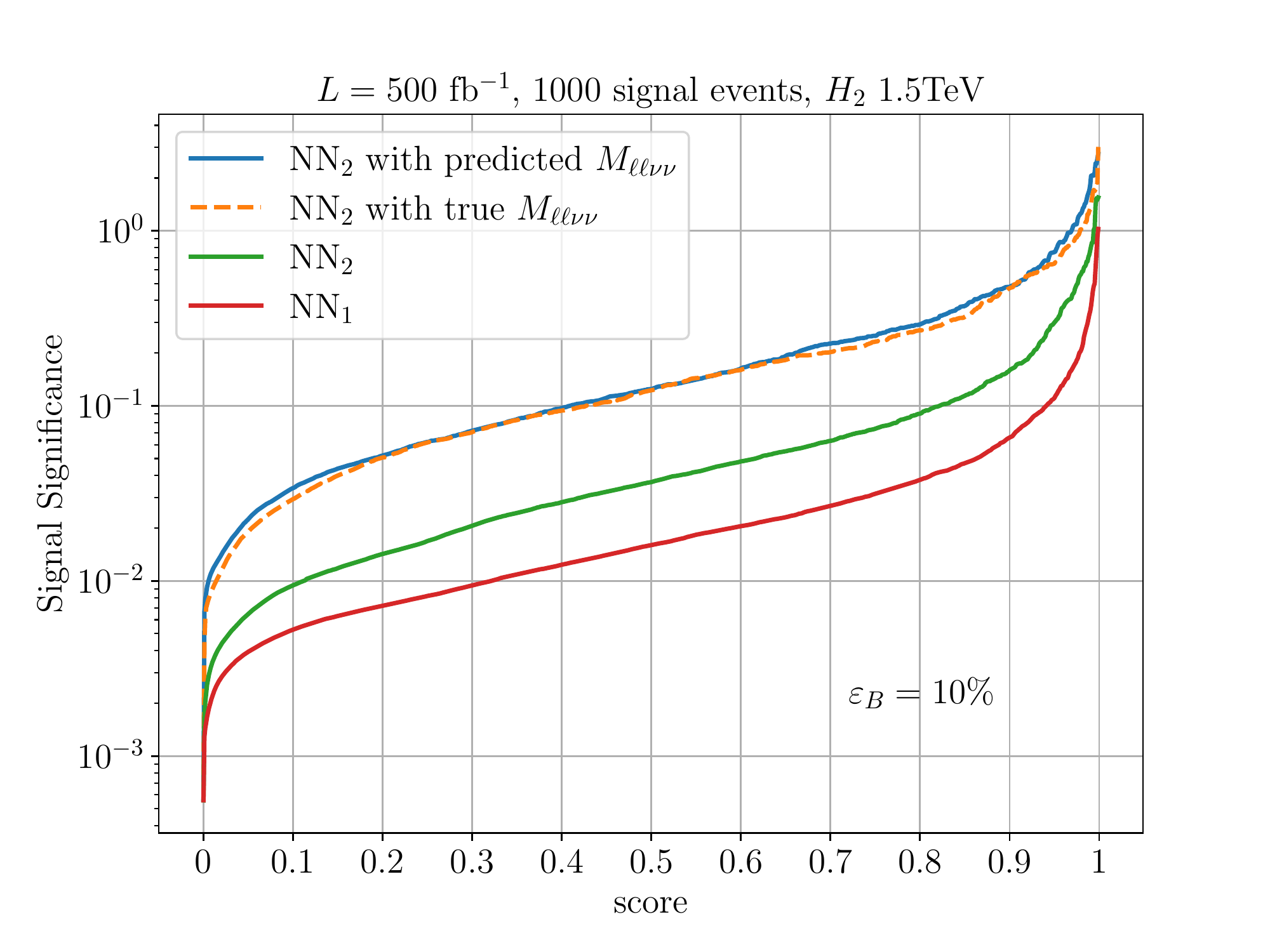}
     \includegraphics[scale=0.35]{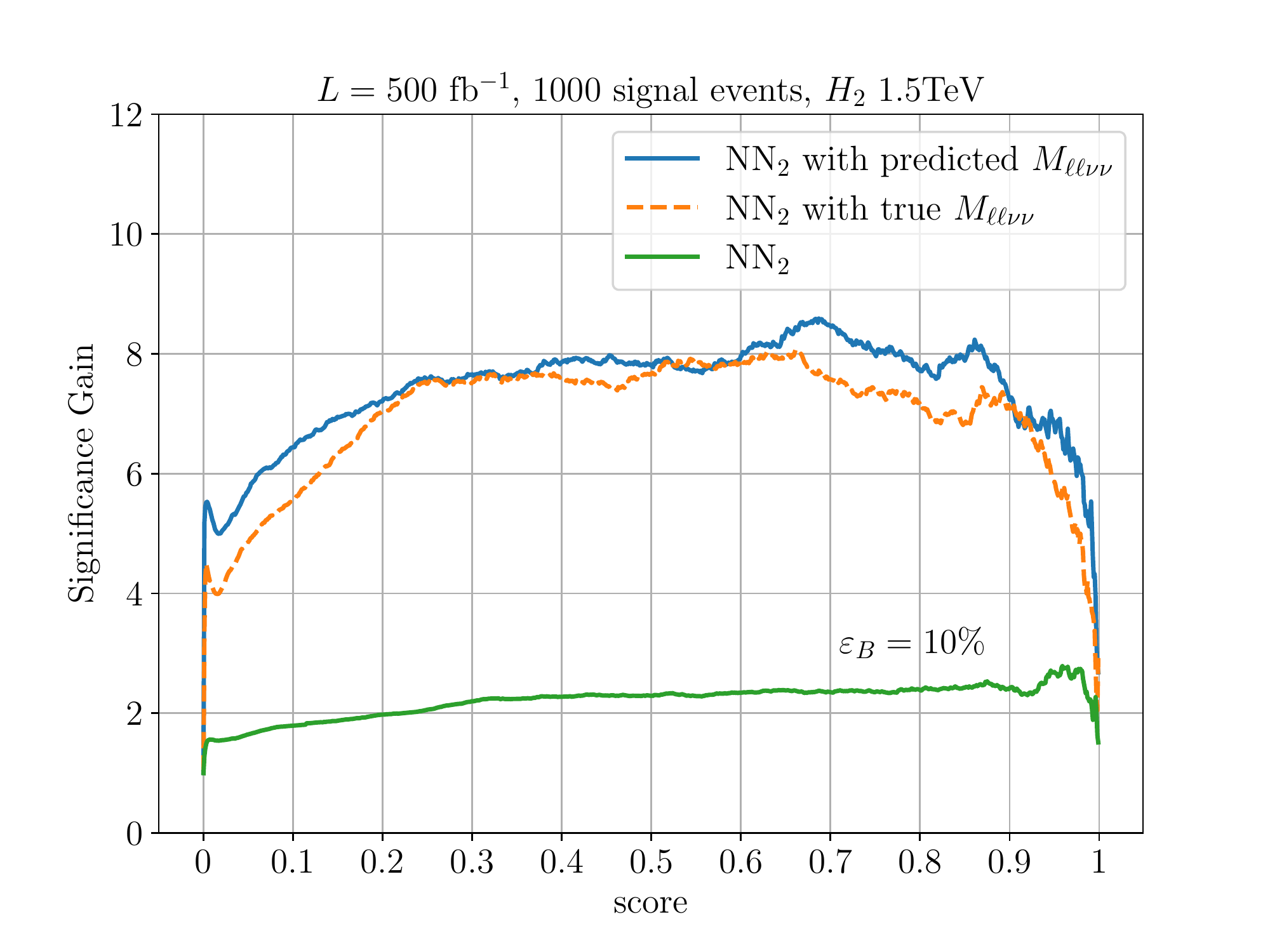}\\
     \includegraphics[scale=0.35]{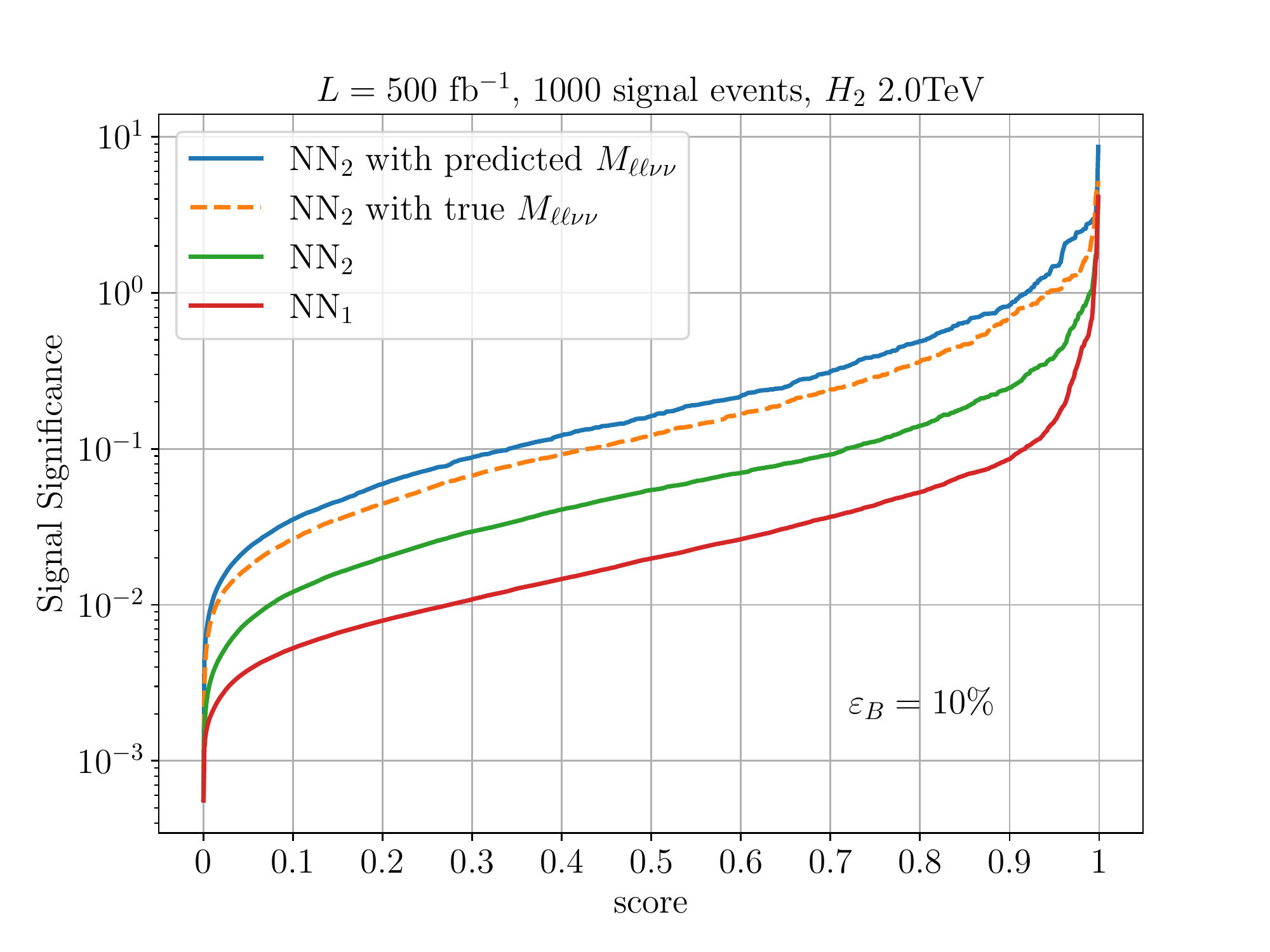}
     \includegraphics[scale=0.35]{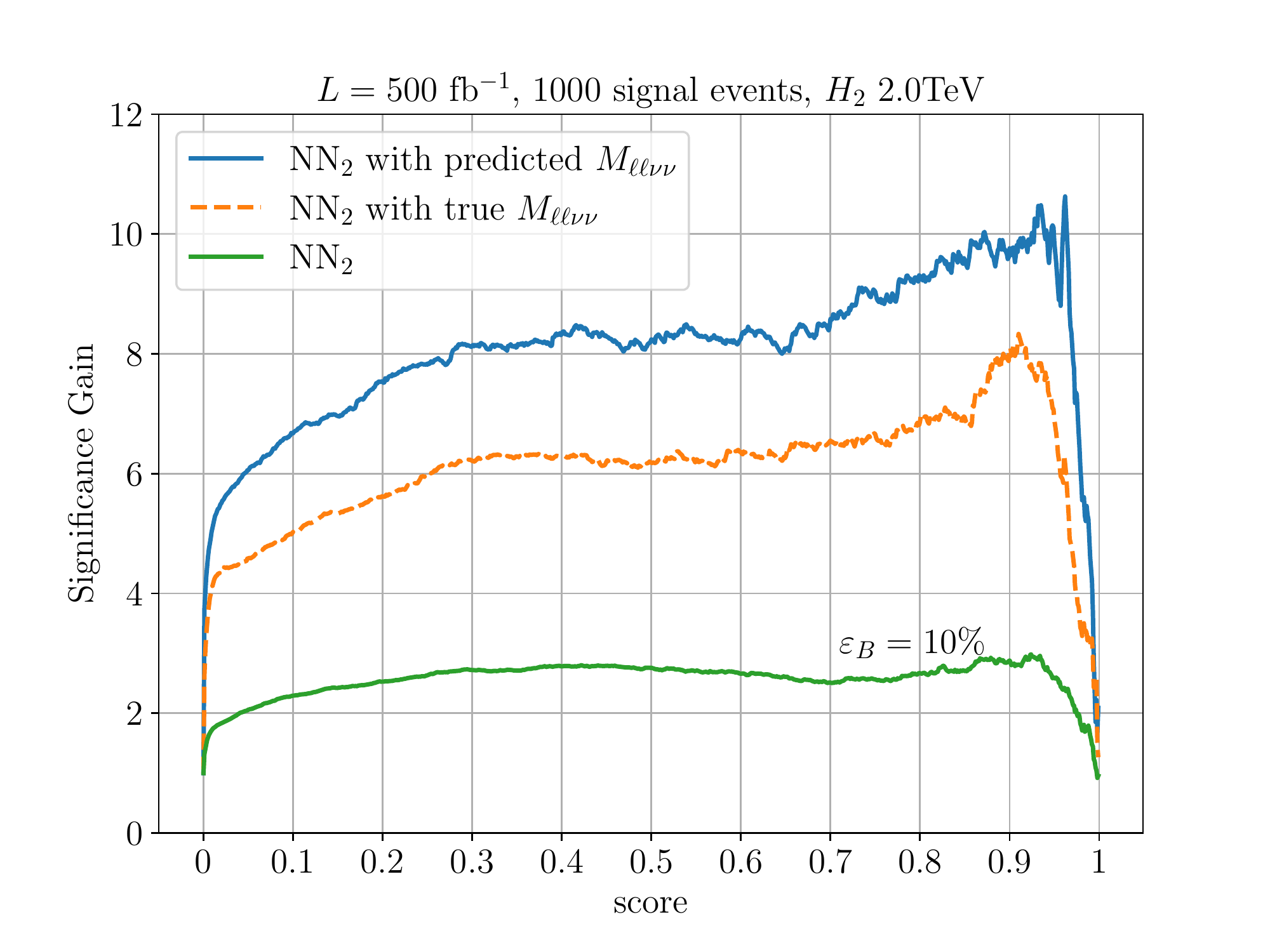}
     \caption{The statistical significance of the $k$NNNN algorithm as a function of the cut on the output score (left panels) and the gain in the significance compared to the NN$_1$ classifier (right panels) for a 1 (upper row), 1.5 (center row) and 2 TeV (lower row) Higgs mass. The systematic uncertainty is fixed at 10\%.}
     \label{fig:results}
 \end{figure}
 In Fig.~\eqref{fig:results}, we show, at the left panels, the statistical significance, assuming a $\varepsilon_B=10$\% systematic uncertainty in the backgrounds rates, for a new Higgs boson of 1, 1.5, and 2 TeV mass, from top to bottom rows, respectively. To raise the significance, we cut on the classifiers' signal score output represented in the plots' horizontal axis. The 1st NN and 2nd NN lines depict the significance of NN$_1$ and NN$_2$, respectively, without including the $\mllvv$ prediction. Even without reconstructing the resonance, the stacking of the neural networks boosts the significance, as expected. The statistical significance is much enhanced, including the predicted mass, as shown in the top lines in all the left panels. As we see from the dashed lines, the agreement with what should be expected using the true masses in the data representation is good. The agreement is better for lower masses, while a more pronounced overestimation is observed in the 2 TeV case. An insufficient number of simulated background samples might cause that effect. Yet, the quality of the resonance reconstruction enables us to employ the method to select the signal events better. 

At the right panels of Fig.~\eqref{fig:results}, we show the significance gain relative to the first neural network classifier, NN$_1$. While not including the predicted $\mllvv$ mass leads to gains around 2, including them boosts the gains to up to 6, 8 and 10 for 1, 1.5 and 2 TeV masses, respectively. As noted in the left panels, there is a more pronounced overestimation of around 20 to 25\%, depending on the cut score, for 2 TeV Higgs bosons. Similar gains were observed when we varied the Higgs bosons widths down to $\Gamma_H/m_H=1$\%. The train/test/validation dataset was randomly split five times to assess the robustness of these results, and tiny variations were observed in this cross-validation.


\section{Conclusions and Prospects}
\label{sec:conclusions}

As the search for new physics intensifies following the LHC program schedule, new ways to identify particles that hide information through invisible decays are surely welcome. In this work, we designed an algorithm capable of reconstructing the mass of a new heavy Higgs boson decaying to $W^+W^-\to \ell^+\ell^{-\prime}\nu_\ell\bar{\nu}_{\ell^\prime}$ and its main SM backgrounds using a simple but adequately tuned nearest neighbors algorithm. The algorithm assumes the previous knowledge of the event classes and the Higgs boson mass; therefore, it is useful for post-discovery studies, for example, an analysis that requires a selection of on-mass shell Higgs bosons.

 More importantly, including the predicted $k$NN $\mllvv$ mass as an attribute for a neural network classification improves the accuracy, the true and false positives/negatives rates, and the likelihood of true class classifications when compared to a neural network that does not have a clue about the masses. The gain in the statistical significance is the ultimate test for the proposed algorithm. We found a gain factor in significance up to a factor of 10 for a 2 TeV Higgs boson mass. For lighter masses, of 1 and 1.5 TeV, the gains are less pronounced but also high, up to 6 and 8, respectively, depending on the cut placed on the signal class score. We checked that the predicted mass is reliable and robust as a new feature for classification by comparing our results against classifiers trained with the true $\mllvv$ masses. Not only the binned invariant mass distributions agree but also the final statistical significance agree within a few tens of percent, at most. 
 
 The $k$NNNN algorithm can be applied to other observable variables as well. For example, the scattering angle of the $W$ bosons can be obtained in the fully leptonic channel beside the charged leptons angles. The masses of particles in different topologies can also be obtained. For example, we guess that sparticles' mass distributions from decay chains of various lengths might be recovered after their determination with other methods. 
 
 The next step in this kind of investigation is to relax the previous knowledge of the mass parameters and weaken the level of supervision when training the classifiers and regressors. 
 Outlier detection and other unsupervised techniques can be readily used to dismiss previous knowledge of the signal class, yet, using $k$NN for regression requires the knowledge of mass parameters. A completely weakly supervised regression algorithm that assumes just the knowledge of the background classes is challenging once it involves generalization across classes with essential information loss. We are currently investigating deep neural networks and variational autoencoders for regression algorithms trained on a single background class but still assuming previous knowledge of signal mass parameters. These results will be presented elsewhere. 
 
 \acknowledgments
 A. Alves is supported by grants from the Conselho Nacional de Desenvolvimento Cient\'{\i}fico e Tecnol\'ogico (CNPq) (307317/2021-8) and Funda\c{c}\~ao de Amparo \`a Pesquisa do Estado de S\~ao Paulo (FAPESP) (2021/01089-1).


\bibliographystyle{apsrev4-1} 
\bibliography{referencias}

\end{document}